\documentclass[11pt]{article}
\usepackage{latexsym}
\usepackage{amsfonts} 
\newtheorem{theorem}{Theorem}
\newtheorem{lemma}{Lemma}

\begin{document}
\title{Global foliations of matter spacetimes with Gowdy symmetry} 
\author{H\aa kan Andr\'{e}asson\\Department of Mathematics\\
Princeton University\\Princeton, NJ 08540\\
hand@math.princeton.edu\footnote{Permanent address: 
Department of Mathematics,
 Chalmers University of Technology, S-412 96 G\"{o}teborg, Sweden. 
email: hand@math.chalmers.se}} 
\date{} 
\maketitle 

\begin{abstract}
A global existence theorem, with respect to a geometrically defined time, is 
shown for Gowdy symmetric globally hyperbolic solutions 
of the Einstein-Vlasov system for arbitrary (in size) initial data. 
The spacetimes being studied contain 
both matter and gravitational waves. 
\end{abstract} 

\section{Introduction} 
An important problem in classical general relativity is the question 
of global existence (in an appropriate sense) for globally hyperbolic 
solutions of the vacuum-Einstein and 
matter-Einstein equations. The main 
motivation being its relationship to the cosmic censorship conjectures. 
Strong cosmic censorship has eg. by Eardley and Moncrief [EM] been formulated 
as a question on 
global existence and asymptotic behaviour of solutions to the Einstein 
equations, suggesting a definite method of analytical attack. 

To begin studying the long-time behaviour of solutions to a complicated 
partial differential equation system one might focus on families of 
solutions with some prescribed symmetry. With the exception 
of the monumental work on global nonlinear stability of the Minkowski 
space by Christodoulou and Klainerman [CK], the practice in general 
relativity has for long been to study ``global existence'' problems 
under symmetric assumptions. 

One family of (cosmological) solutions which have been studied extensively 
is the Gowdy spacetimes [G]. 
These spacetimes are vacuum but admit gravitational waves (in contrast
to eg. spherically symmetric spacetimes). 
Global existence has been 
shown for the Gowdy spacetimes [M], strong cosmic censorship is 
settled in the case of polarized Gowdy spacetimes [CIM], and much is known 
about the subset of the Gowdy spacetimes which admit an extension across a 
Cauchy horizon [CI]. 

In this paper we show global 
existence, with respect to a geometrically defined time, for matter 
spacetimes (Einstein-Vlasov) with Gowdy symmetry 
and thereby we extend Moncrief's result [M] in the vacuum case. This 
is the first result which provides a global foliation of a spacetime 
containing both matter and gravitational waves. Moreover, for matter 
spacetimes there are only a few global results available all together. 
Let us briefly mention some of these results. First, by matter spacetimes 
we have in mind spacetimes where the matter consists of massive particles. 
One can also consider spacetimes which only contains radiation and 
important results have been obtained in this direction, e.g. 
Christodoulou has obtained strong results in the spherically 
symmetric case with a scalar field as matter model (see e.g. [Cu1], [Cu2] 
and the references therein). 
For spacetimes containing massive particles the main global results can 
be summarized as follows. 
Under a smallness condition on the initial data, Rein and Rendall have 
[RR] shown that solutions of the spherically 
symmetric Einstein-Vlasov system are geodesically complete. 
Some information on the large data problem was then obtained in [RRS]. 
Christodoulou has in a series of papers (see [Cu3] and the references 
therein) 
studied the Einstein-Euler equation 
in the spherically symmetric case for a special equation 
of state, adapted to understand the dynamics of 
a supernova explosion. He can globally control the solutions 
to the Cauchy problem and he finds solutions whose behaviour 
resembles qualitatively 
that of a supernova explosion. Finally, the most relevant results 
in the context of this paper are those on cosmological solutions 
by Rendall [Rl1-2] and Rein [Rn]. These are discussed 
in some detail in relation to our result below. 

Our method of proof is inspired by a recent global foliation result for vacuum 
spacetimes admitting a $T^2$ isometry group, acting on $T^3$ 
spacelike surfaces [BCIM]. These spacetimes 
are more general than the Gowdy spacetimes: both families admit 
two commuting Killing vectors 
but in the Gowdy case there is the additional condition that the twists 
are zero. The twists are defined by 
\begin{equation}
c_1=\epsilon_{\mu\nu\rho\delta}X^{\mu}Y^{\nu}\nabla^{\rho}X^{\delta},\;\; 
c_2=\epsilon_{\mu\nu\rho\delta}X^{\mu}Y^{\nu}\nabla^{\rho}Y^{\delta},
\label{twist} 
\end{equation} 
where $X,Y$ are Killing vectors associated with the isometry group. 
It follows from the Einstein equations that in vacuum these quantities 
are constant throughout spacetime [G]. 

One difficulty in studying long-time existence problems in general relativity 
is the lack of having a fixed time measure. 
A solution which remains regular for an infinite 
range of one time scale may become singular within a finite range of another. 
In [BCIM] this problem is treated by choosing a coordinate system 
in which the time is fixed to the geometry of spacetime. In fact, the time is 
defined to be the area of the two dimensional spacelike orbits of the $T^2$ 
isometry group. These coordinates are called areal coordinates. 
The main theorem in [BCIM] shows that the entire maximal globally 
hyperbolic development of the initial hypersurface can be foliated by 
areal coordinates. These coordinates are however only used in a direct way 
in the future direction. To show that the past of the initial 
hypersurface is 
covered by areal coordinates the authors use conformal coordinates 
(the time is not fixed to the geometry of spacetime) 
in which the equations take a more suitable form for an analytical treatment. 
By a long chain of geometrical arguments it is then shown that the development 
in conformal coordinates admits a 
foliation by areal coordinates, and that it covers the past maximal globally 
hyperbolic 
development of the initial hypersurface. 

We prove that $T^3\times\mathbb{R}-$matter spacetimes with Gowdy 
symmetry admit 
global foliations by areal coordinates. The matter content is described 
by the Vlasov equation. This is a kinetic equation and gives a statistical 
description 
of a collection of collisionless ``particles''. In the cosmological case 
the particles are galaxies or clusters of galaxies 
whereas in stellar dynamics they are stars. The Vlasov equation has shown 
to be suitable in general relativity for the study of the long-time 
behaviour of 
matter in gravitational fields. In particular it rules out the formation of 
shell-crossing singularities. For a discussion on the choice of matter 
model see [Rl4] and [Rl5].  

To prove the existence of a global folitaion we also work directly 
in areal coordinates in the expanding (future) direction, and 
in the contracting (past) direction, 
we first show a global existence theorem in conformal coordinates and 
then we invoke the geometrical arguments in [BCIM] to complete the proof. 
We point out that our result depends strongly on the exact strucure of 
the Vlasov equation 
and do not hold for general matter models which are only restricted by 
certain inequalities on the components of the energy-momentum tensor. 
 
A related and interesting result has recently been shown by Rendall 
[Rl1] (see also [Rl2]). He 
considers $T^2$ symmetric spacetimes for the Einstein-Vlasov and 
the Einstein-wave map equations and he shows that if such a spacetime admits 
at least one compact constant mean curvature (CMC) hypersurface then the past 
of that surface can be covered by a foliation of compact CMC hypersurfaces. 
The CMC- and the areal coordinate foliation are both geometrically 
based time foliations which provide frameworks for studying 
strong cosmic censorship and other global issues. 
The main motivation for developing techniques to obtain CMC foliations 
is that the definition of 
a CMC hypersurface does not depend on any symmetry assumptions and it is 
hence possible that CMC foliations will exist for rather general spacetimes. 
The areal coordinate foliation used here is less general since 
it is adapted to the symmetry, but leads in the Gowdy 
case (note that the results in [Rl1] apply to the more general $T^2$ symmetric 
spacetimes which we hope to explore in the future) 
to stronger results. 
Namely, the arguments in [Rl1] do not show that the entire future 
of the initial 
hypersurface can be covered, and the existence of the CMC foliation is only 
guaranteed under the hypothesis that spacetime admits at least one 
such hypersurface. 

We also mention a result in this direction due to 
Rein [Rn]. He 
has studied cosmological Einstein-Vlasov spacetimes with stronger symmetry 
restrictions than in the Gowdy 
case (the spacetimes admit three Killing vectors). 
In these spacetimes gravitational waves cannot exist. 
For plane symmetry (the relevant case for us) he has shown existence back 
to the initial singularity for 
small initial data, and under the assumption that 
one of the field components is bounded, 
he obtains global 
existence for large data in the future direction. An interesting 
result in his work is that the initial singularity is shown to be a 
curvature singularity as well as a ``crushing'' singularity (see [ES]). 

The outline of the paper follows in the large that of [BCIM]. In section 2 we 
describe Gowdy symmetry and give the equations for the Einstein-Vlasov system 
in areal and conformal coordinates. The main theorem is formulated 
in section 3 where we also describe the geometrical arguments in [BCIM] 
needed to complete the proof in the contracting direction. 
Section 4 is devoted to the analysis in the contracting direction. 
Estimates for the field components and the matter terms are derived 
in conformal coordinates, by using e.g. light-cone arguments and methods 
originally developed for the Vlasov-Maxwell equation. 
The analysis in the expanding direction is carried out in 
areal coordinates in section 5 where a number of estimates are derived. 
Light-cone arguments and an ``energy'' monotonicity lemma are important tools 
for obtaining bounds on the field components and their derivatives. 
The control of the matter terms and their derivatives rely on 
three lemmas. The first one is the ``energy'' monotonicity lemma just 
mentioned. Then, in the second lemma a careful analysis of the characteristic 
system associated with the Vlasov equation is carried out, which leads to 
a bound on the support of the momenta. 
The third lemma provides bounds on the derivatives of the matter 
terms and relies indirectly on the geodesic deviation equation. 
This equation relates the curvature tensor and the acceleration 
of nearby geodesics and has proved useful in previous studies 
of the Einstein-Vlasov system (see [RR], [Rn] and [Rl3]). 

\section{The Einstein-Vlasov system with Gowdy symmetry}
Let us begin with a brief review of Gowdy symmetry. Consider a spacetime 
that can be foliated by a family of compact, connected, and orientable 
hypersurfaces. If the maximal isometry group of the spacetime is two 
dimensional, and if it acts invariantly and effectively on the foliation, 
then the isometry group must be $U(1)\times U(1)$.   
Moreover, the foliation surfaces must be homeomorphic to $T^{3}, S^1\times 
S^2, S^3$ or $L(p,q)$ (the Lens space), and the action is unique up to 
equivalence. The Killing vector fields $X,\; Y$ associated with the isometry 
group have to commute in such a spacetime. 
We say that spacetimes satisfying the symmetry 
conditions above and in which both the twists $c_1, c_2$ (see \ref{twist}) 
vanish have 
Gowdy symmetry. We remark that the term ``Gowdy spacetime'' is reserved for 
the vacuum case. For more background on Gowdy symmetry we refer 
to [G], [Cl]. 

As mentioned above there are several choices of spacetime manifolds
compatible with Gowdy symmetry. In this paper we restrict our attention
to the $T^{3}-$case. It is an interesting fact that in vacuum this is 
the only
possibility if the condition of vanishing twists is relaxed. 

The dynamics of the matter is governed by the Vlasov equation. This 
is a kinetic equation and models a collisionless system of particles, 
i.e. the particles follow the geodesics of spacetime. 
For a nice introduction to the Einstein-Vlasov system see [Rl3]. 
We also mention the survey of Ehlers [E] for more information 
on kinetic theory in general relativity, and the book by Binney 
and Tremaine [BT] for some applications of kinetic theory 
in stellar dynamics. 

We will use two choices of coordinates, areal coordinates and
conformal coordinates. It has been shown in [Cl] that, at least
locally, any globally hyperbolic (non-flat) Gowdy spacetime 
on $T^{3}\times\mathbb{R}$
admits each of these coordinates. Both sets of coordinates are chosen
so that $$X=a\frac{\partial}{\partial x}+b\frac{\partial}{\partial
  y},$$ and $$Y=c\frac{\partial}{\partial x}+d\frac{\partial}{\partial y}$$ 
are Killing vector fields ($a,b,c$ and $d$ are constants with
$ad-bc\not=0$), and in both cases $\theta\in S^{1}$ denotes the remaining 
spatial coordinate. Below the form of the metric and the Einstein-Vlasov
system is given in areal and conformal coordinates. The functions 
$R,\alpha,U,A,\eta$ all depend on $t$ and $\theta$ and 
the function $f$ depends on $t,\theta$ and $v\in\mathbb{R}^3$.\\ 
\begin{center}
\textbf{Areal Coordinates}
\end{center} 
\vspace{.3cm}
Metric 
\begin{equation} 
g=-\mbox{e}^{2(\eta-U)}\alpha dt^{2}+\mbox{e}^{2(\eta-U)}
d\theta^{2}+\mbox{e}^{2U}(dx+Ady)^{2}+
\mbox{e}^{-2U}t^{2}dy^{2}\label{areal} 
\end{equation} 
\phantom{hej}\\
\vspace{.3cm}
The Einstein-matter constraint equations
\begin{eqnarray}
\displaystyle\frac{\eta_{t}}{t}&=&U_{t}^{2}+\alpha U_{\theta}^{2}+
\frac{\mbox{e}^{4U}}{4t^2}(A_{t}^2+\alpha A_{\theta}^2)+
\mbox{e}^{2(\eta-U)}\alpha\rho\label{constr1}\\
\displaystyle\frac{\eta_{\theta}}{t}&=&2U_{t}U_{\theta}+
\frac{\mbox{e}^{4U}}{2t^2}
A_{t}A_{\theta}-\frac{\alpha_{\theta}}{2t\alpha}-\mbox{e}^{2(\eta-U)}\sqrt{\alpha}J\label{constr2}\\
\displaystyle\alpha_{t}&=&2t\alpha^{2}\mbox{e}^{2(\eta-U)}(P_{1}-\rho)\label{constr3}
\end{eqnarray} 
\phantom{hej}\\
\vspace{.3cm}
The Einstein-matter evolution equations 
\begin{eqnarray} 
\displaystyle\eta_{tt}-\alpha\eta_{\theta\theta}&=&\frac{\eta_{\theta}\alpha_{\theta}}{2}+
\frac{\eta_{t}\alpha_{t}}{2\alpha}-\frac{\alpha_{\theta}^{2}}{4\alpha}+\frac{\alpha_{\theta\theta}}{2}-U_{t}^{2}+\alpha
U_{\theta}^{2}+\frac{\mbox{e}^{4U}}{4t^2}(A_{t}^2-\alpha A_{\theta}^2)
\nonumber\\
\displaystyle & &-\alpha\mbox{e}^{2(\eta-U)}P_{3}-\frac{A^2}{t^2}\alpha
\mbox{e}^{2(\eta+U)}P_2-\frac{2A}{t}\alpha\mbox{e}^{2\eta}S_{23}
\label{evol1}\\ 
\displaystyle U_{tt}-\alpha U_{\theta\theta}&=&-\frac{U_{t}}{t}+\frac{U_{\theta}\alpha_{\theta}}{2}+\frac{U_{t}\alpha_{t}}{2\alpha}+
\frac{\mbox{e}^{4U}}{2t^2}(A_{t}^2-\alpha A_{\theta}^2)\nonumber\\
\displaystyle & &
+\frac{1}{2}\mbox{e}^{2(\eta-U)}\alpha(\rho-P_{1}+P_{2}-P_{3})\label{evol2}\\
\displaystyle A_{tt}-\alpha A_{\theta\theta}&=&\frac{A_t}{t}+
\frac{\alpha_{\theta}A_{\theta}}{2}+\frac{\alpha_{t}A_{t}}{2\alpha}-4A_tU_t
+4\alpha A_{\theta}U_{\theta}
\nonumber\\ 
\displaystyle & &+2t\alpha\mbox{e}^{2(\eta-2U)}S_{23}\label{evol4} 
\end{eqnarray} 
\phantom{hej}\\
\vspace{.3cm}
The Vlasov equation
\begin{eqnarray}
&\displaystyle\frac{\partial f}{\partial
  t}+\frac{\sqrt{\alpha}v^{1}}{v^{0}}\frac{\partial
  f}{\partial\theta}-\left[
(\eta_{\theta}-U_{\theta}+\frac{\alpha_{\theta}}{2\alpha})\sqrt{\alpha}v^{0}+(\eta_{t}-U_{t})v^{1}-\frac{\sqrt{\alpha}\mbox{e}^{2U}A_{\theta}}{t}
\frac{v^2v^3}{v^0}
\right.& 
\nonumber\\
&\displaystyle\left.
+\frac{\sqrt{\alpha}U_{\theta}}{v^{0}}((v^{3})^{2}-(v^{2})^{2})\right]\frac{\partial
  f}{\partial v^{1}} 
-\left[U_{t}v^{2}+\sqrt{\alpha}U_{\theta}\frac{v^{1}v^{2}}{v^{0}}\right]
\frac{\partial f}{\partial v^{2}}& 
\nonumber\\
&\displaystyle-\left[(\frac{1}{t}-U_{t})v^{3}-\sqrt{\alpha}U_{\theta}\frac{v^{1}v^{3}}{v^{0}}+\frac{\mbox{e}^{2U}v^2}{t}
(A_t+\sqrt{\alpha}A_{\theta}\frac{v^1}{v^0})\right]\frac{\partial
  f}{\partial v^{3}}=0&\label{vv} 
\end{eqnarray} 
\phantom{hej}\\
The matter quantities
\begin{eqnarray} 
\rho(t,\theta)&=&\int_{\mathbb{R}^{3}}v^{0}f(t,\theta,v)\;dv\label{matt1}\\ 
P_{k}(t,\theta)&=&\int_{\mathbb{R}^{3}}\frac{(v^{k})^{2}}{v^{0}}f(t,\theta,v)\;dv,\;\;\;k=1,2,3\label{matt2}\\
J(t,\theta)&=&\int_{\mathbb{R}^{3}}v^{1}f(t,\theta,v)\;dv\label{matt3}\\ 
S_{23}(t,\theta)&=&\int_{\mathbb{R}^{3}}\frac{v^2v^3}{v^0}f(t,\theta,v)\;dv
\label{matt4} 
\end{eqnarray}
Here the variables $v$ are related to the canonical momenta $p$ through
\begin{equation}
v^{0}=\sqrt{\alpha}\mbox{e}^{\eta-U}p^{0},\;v^{1}=\mbox{e}^{(\eta-U)}p^{1},\;
v^{2}=\mbox{e}^{U}p^{2}+A\mbox{e}^{U}p^{3},\;
v^{3}=t\mbox{e}^{-U}p^{3},\label{pmu} 
\end{equation}
and 
$$p^{\mu}:=\frac{dx^{\mu}}{d\tau},\;\;x^{\mu}=(t,\theta,x,y),$$ where 
$\tau$ is proper time. It is assumed that all 
'particles' have the same mass (normalized to one) and follow the
geodesics 
of spacetime (collisionless particle system). 
Hence $$g_{\mu\nu}p^{\mu}p^{\nu}=-1,$$ 
so that 
\begin{equation} 
v^{0}=\sqrt{1+(v^{1})^{2}+(v^{2})^{2}+(v^{3})^{2}}.\label{v0} 
\end{equation} 

In conformal coordinates the function $\alpha$ is removed, having 
the consequence that the orbital area function $R$ now depends on 
both $t$ and $\theta$ (in areal coordinates $R=t$). 
In these coordinates the metric and the Einstein-Vlasov system take
the following form. 
\begin{center}
\textbf{Conformal coordinates} 
\end{center}
\vspace{.3cm}
Metric
\begin{equation}
g=\mbox{e}^{2(\eta-U)}(-dt^{2}+d\theta^{2})+\mbox{e}^{2U}(dx+Ady)^{2}+
\mbox{e}^{-2U}R^{2}dy^{2}\label{confl} 
\end{equation} 
\phantom{hej}\\
\vspace{.3cm}
The Einstein-matter constraint equations
\begin{eqnarray}
&\displaystyle
U_{t}^{2}+U_{\theta}^{2}+\frac{\mbox{e}^{4U}}{4R^2}(A_{t}^2+A_{\theta}^2)
+\frac{R_{\theta\theta}}{R}-\frac{\eta_{t}R_{t}}{R}-\frac{\eta_{\theta}R_{\theta}}{R}=-\mbox{e}^{2(\eta-U)}\rho&\label{constr1con}\\
&\displaystyle
2U_{t}U_{\theta}+\frac{\mbox{e}^{4U}}{2R^2}A_{t}A_{\theta}
+\frac{R_{t\theta}}{R}-\frac{\eta_{t}R_{\theta}}{R}
-\frac{\eta_{\theta}R_{t}}{R}=\mbox{e}^{2(\eta-U)}J&\label{constr2con}\\ 
\end{eqnarray} 
\phantom{hej}\\
\vspace{.2cm}
The Einstein-matter evolution equations
\begin{eqnarray} 
\displaystyle U_{tt}-U_{\theta\theta}&=&\frac{U_{\theta}R_{\theta}}{R}-\frac{U_{t}R_{t}}{R}+\frac{\mbox{e}^{4U}}{2R^2}(A_{t}^2-A_{\theta}^2)
\nonumber\\ 
\displaystyle & &+\frac{1}{2}\mbox{e}^{2(\eta-U)}(\rho-P_{1}+P_{2}-P_{3})\label{evol1con}\\ 
\displaystyle A_{tt}-A_{\theta\theta}&=&\frac{R_tA_t}{R}-\frac{R_{\theta}A_{\theta}}{R}
+4(A_{\theta}U_{\theta}-A_tU_t)+2R\mbox{e}^{2(\eta-2U)}S_{23}\label{evol4con}
\\ 
\displaystyle R_{tt}-R_{\theta\theta}&=&R\mbox{e}^{2(\eta-U)}(\rho-P_{1})
\label{evol2con}\\ 
\displaystyle\eta_{tt}-\eta_{\theta\theta}&=& 
U_{\theta}^{2}-U_{t}^{2}+\frac{\mbox{e}^{4U}}{4R^2}(A_{t}^2-A_{\theta}^2)
-\mbox{e}^{2(\eta-U)}P_{3}
\nonumber\\ 
\displaystyle & &-\frac{A^2}{R^2}\mbox{e}^{2(\eta+U)}P_2-\frac{2A}{R}\mbox{e}^{2\eta}S_{23}\label{evol3con} 
\end{eqnarray}
\phantom{hej}\\
\vspace{.2cm}
The Vlasov equation
\begin{eqnarray}
&\displaystyle\frac{\partial f}{\partial
  t}+\frac{v^{1}}{v^{0}}\frac{\partial
  f}{\partial\theta}-\left[
(\eta_{\theta}-U_{\theta})v^{0}
+(\eta_{t}-U_{t})v^{1}-U_{\theta}\frac{(v^{2})^{2}}{v^{0}}\right.& 
\nonumber\\ 
&\displaystyle\left. +(U_{\theta}-\frac{R_{\theta}}{R})
\frac{(v^{3})^{2}}{v^{0}}-\frac{A_{\theta}}{R}\mbox{e}^{2U}
\frac{v^2v^3}{v^0}\right]
\frac{\partial f}{\partial v^{1}}
-\left[U_{t}v^{2}+U_{\theta}\frac{v^{1}v^{2}}{v^{0}}
\right]\frac{\partial f}{\partial v^{2}}&
\nonumber\\ 
&\displaystyle -\left[(\frac{R_{t}}{R}-U_{t})v^{3}-(U_{\theta}
-\frac{R_{\theta}}{R})\frac{v^{1}v^{3}}{v^{0}}
+\frac{\mbox{e}^{2U}v^2}{R}(A_t+A_{\theta}\frac{v^1}{v^0})
\right]\frac{\partial f}{\partial v^{3}}=0&\label{vvcon} 
\end{eqnarray} 
The matter quantities $\rho, P_{k},J$ and $S_{23}$ are given by
(\ref{matt1})-(\ref{matt4}), where in this case 
\begin{equation} 
v^{0}=\mbox{e}^{\eta-U}p^{0},\;v^{1}=\mbox{e}^{(\eta-U)}p^{1},\;
v^{2}=\mbox{e}^{U}p^{2}+A\mbox{e}^{U}p^{3},\;v^{3}=R\mbox{e}^{-U}p^{3}, 
\end{equation}
and (\ref{v0}) holds here as well. 

\textbf{Remark. }It might be instructive to relate the metric in (\ref{confl}) 
with that used by Rein [Rn] mentioned in the introduction. By letting $A=0$ 
and $U=(1/2)\ln{R}$ in (\ref{confl}) we obtain a metric which admits three 
Killing vectors and which depends on two field components. The distribution 
function $f$ depends in this case on $p^1$ and $(p^2)^2+(p^3)^2.$ 

\section{The main theorem} 
Let $(h,k,f_0)$ be a Gowdy symmetric initial data set on $T^3$, by 
this we mean 
that $h$ is a Riemannian metric on $T^3$, invariant under an effective $T^2$ 
action; $k$ is a symmetric 2-tensor on $T^3$, also invariant under the same 
$T^2$ group action; the twists $c_1$ and $c_2$ are both zero; the initial 
distribution function $f_0$ is defined on $T^3$ and is invariant under 
the same $T^2$ group action and possesses the following additional symmetry, 
which reads, in coordinates that cast the metric in the forms (\ref{areal}) or 
(\ref{confl}), $f_0(\theta,p^1,p^2,p^3)=f_0(\theta,p^1,-p^2,p^3)=
f_0(\theta,p^1,p^2,-p^3)$ (this assumption is necassary for 
the Einstein-Vlasov system to be compatible with the form of the metric); 
and that $(h,k,f_0)$ satisfy the Einstein-Vlasov constraint equations. 
We also assume that $(h,k)$ are $C^\infty$ on $T^3$ and that $f_0$ is 
a non-negative, not identically zero, $C^\infty$ function of compact support 
on the tangent bundle $T(T^3)$ of $T^3$. 

\textbf{Remark. }The smoothness assumption on the initial data 
is not a necassary condition. It is included so that we can refer 
directly to the classical local existence theorems. However, 
the estimates in this paper provide the information needed 
for proving a local existence theorem for $C^2\times C^1$ data 
$(h,k)$ and $C^1$ data $f_0$. Moreover, the assumption $f_0\not= 0$ is here 
included for a technical reason and we refer to [M] or section 5 in 
this paper for the vacuum case. Indeed, it 
is in this case possible to work directly in areal coordinates and 
the estimates derived in section 5 are sufficient. See also the remark 
following lemma 1 in that section.\\ 

The results by Choquet-Bruhat [CB] and Choquet-Bruhat and 
Geroch [CBG], show that there exists 
a unique maximal globally hyperbolic development $(\Sigma\times\mathbb{R},g,f)$ of a 
given initial data set on 
a three-dimensional manifold $\Sigma$ for the Einstein-Vlasov equation. 
Let us briefly comment upon the initial conditions imposed. 
The relations between a given initial data set 
$(h,k)$ on a three-dimensional manifold $\Sigma$ and the metric $g$ 
on the spacetime manifold is that there exists an imbedding $\psi$ of $\Sigma$ 
into the spacetime such that the induced metric and second fundamental form of 
$\psi(\Sigma)$ coincide with the result of transporting $(h,k)$ with $\psi$. 
For the relation of the distribution functions $f$ and $f_0$ we have to 
note that $f$ is defined on the mass shell 
(for $m=1$ it is the set of all future pointing
unit timelike vectors). The initial condition imposed is that
the restriction of $f$ to the part of the mass shell over $\psi(\Sigma)$ 
should be equal to $f_0\circ (\psi^{-1},d(\psi)^{-1})\circ\phi$ where $\phi$ 
sends each point of the mass shell over $\psi(\Sigma)$, to its orthogonal 
projection onto the tangent space to $\psi(\Sigma)$. 

Our main theorem now reads, 
\begin{theorem} 
Let $(h,k,f_0)$ be a smooth Gowdy symmetric initial data set on $T^3$. 
For some non-negative constant $c$, there exists a globally hyperbolic 
spacetime $(\mathcal{M},g,f)$ such that\\ 
(i) $\mathcal{M}=(c,\infty)\times T^3$\\ 
(ii) $g$ and $f$ satisfy the Einstein-Vlasov equation\\ 
(iii) $\mathcal{M}$ is covered by areal coordinates $(t,\theta,x,y)$, with 
$t\in(c,\infty)$, so the metric globally takes the form (\ref{areal}).\\ 
(iv) $(\mathcal{M},g,f)$ is isometrically diffeomorphic to the maximal globally 
hyperbolic development of the initial data $(h,k,f_0)$. 
\end{theorem} 

As described in the introduction we prove global existence in 
conformal and areal coordinates for the past and future directions 
respectively. Then, in order to prove Theorem 1 in the past 
direction, we need to invoke 
substantial geometrical arguments from [BCIM]. 
For the future direction only a simple geometrical argument is 
needed for completing the proof. 
It should be pointed 
out that even if the geometrical results in [BCIM] 
concern the vacuum case they are true also for matter spacetimes 
as long as the Einstein-matter equations form a well-posed 
hyperbolic system, which of course is the case here. 

In section 4 we show that the past maximal development 
of $(h,k,f_0)$ in terms of 
conformal coordinates, which we denote by $D^{-}_{conf}(h,k,f_0)$, has 
$t\rightarrow -\infty$ as long as $R$ stays bounded away from zero. 
Starting from this result we briefly describe how the geometrical 
arguments in [BCIM] lead to a proof of Theorem 1 in the past direction. 

First, in [BCIM] $R$ is shown to be positive everywhere in the globally 
hyperbolic 
region of 
a $T^2$ symmetric spacetime. Also, along any past inextendible timelike path 
in $D^{-}_{conf}(h,k,f_0)$, $R$ is shown to approach a 
limit $R_0\geq 0$ (to be identified with $c$ in Theorem 1), which is 
independent of 
the choice of path. Moreover, for any $\tilde{R}\in (R_0,R_1)$, where $R_1$ 
is the minimum value of $R$ on the initial hypersurface, the level set 
$R=\tilde{R}$ in $D^{-}_{conf}(h,k,f_0)$ is shown to be a Cauchy surface. 
From these facts 
it follows from arguments in [Cl] that $D^{-}_{conf}(h,k,f_0)$ admits areal 
coordinates to the past of the hypersurface $R=R_1$. 
Proposition 4 and 5 in [BCIM] then show that 
$D^{-}_{conf}(h,k,f_0)$ is also isometrically diffeomorphic to the maximal 
past development, $D^{-}(h,k,f_0)$ of $(h,k,f_0)$ on $T^3$. 

In the future direction, global existence in 
areal coordinates is almost sufficient for proving Theorem 1. 
The only statement that remains to be proved in Theorem 1 is 
that the future maximal development is covered by $D^{+}_{areal}(h,k,f_0)$. 
This follows from a very short geometrical argument given in the proof 
of Proposition 5 in [BCIM]. 

\section{Analysis in the contracting direction} 
The local existence theorem of Choquet-Bruhat [CB] together with the result of 
Chrusciel (lemma 4.2 in [Cl]) imply that for any Gowdy symmetric initial data 
set $(h,k,f_0)$ on $T^3$, we can find an interval $(\hat{t}_1,\hat{t}_2)$ 
and $C^\infty$ 
functions 
$R,U,\eta$ on $(\hat{t}_1,\hat{t}_2)\times T^3$, and a non-negative $C^\infty$ function 
$f$ on $(\hat{t}_1,\hat{t}_2)\times P$ ($P$ denotes the mass shell) such that: 
these functions 
satisfy the Einstein-Vlasov equations in conformal coordinate form 
and for some $t_0\in (\hat{t}_1,\hat{t}_2)$, the metric $g$ induces initial data on 
the $t_0$-hypersurface which is smoothly spatially diffeomorphic to $(h,k)$,
and the relation between $f$ and $f_0$ given above holds. 

Now, in order to show that $D^{-}_{conf}(h,k,f_0)$ has $t\rightarrow -\infty$, 
as long as $R$ stays bounded away from zero, it is sufficient to 
prove that on any finite time interval $(\tilde{t},t_0]$, 
the functions $R,U,A,\eta,f$ and all their derivatives are uniformly bounded 
and that the supremum of the support of momenta at time $t$, 
\begin{equation}
Q(t):=\mbox{sup}\{|v|:\exists (s,\theta)\in [t,t_0]\times 
S^1\mbox{such that} f(s,\theta,v)\not= 0\}, 
\end{equation} 
is uniformly bounded. 
Note that the last condition implies that the matter quantities and 
their derivatives are 
uniformly bounded (if $|\partial f/\partial x^{\mu}|<C$). 

\textit{Step 1. }(Monotonicity of $R$ and bounds on its first derivatives.)\\ 
This is a key step and relies on Theorem 4.1 in [Cl] together with 
the arguments in [BCIM]. We have to check that the matter terms 
have the right signs so that these arguments still hold. The bounds 
on $R$ and its first derivatives will play a crucial role when we 
control the matter terms below. 

First we show that $\nabla R$ is timelike. 
Let us introduce the null vector fields 
\begin{equation}
\partial_{\xi}=\frac{1}{\sqrt{2}}(\partial_{t}+\partial_{\theta}),\;\;\;
\partial_{\lambda}=\frac{1}{\sqrt{2}}(\partial_{t}-\partial_{\theta)},
\label{null}
\end{equation}
and let us set $F_{\xi}=\partial_{\xi}F,\;F_{\lambda}=\partial_{\lambda}F$ 
for a function $F$. After some algebra it follows that the constraint 
equations (\ref{constr1con}) and (\ref{constr2con}) can be written 
\begin{eqnarray}
&\partial_{\theta}R_{\xi}=\eta_{\xi}R_{\xi}-RU_{\xi}^2-
R\mbox{e}^{2(\eta-U)}(\rho-J),&\label{Rxi}\\ 
&\partial_{\theta}R_{\lambda}=\eta_{\lambda}R_{\lambda}-RU_{\lambda}^2
-R\mbox{e}^{2(\eta-U)}(\rho+J).\label{Rla}& 
\end{eqnarray} 
Let $h_{1}$ and $h_{2}$ be defined by 
$$h_{1}:=RU_{\xi}^2+R\mbox{e}^{2(\eta-U)}(\rho-J),$$ and 
$$h_{2}:=RU_{\lambda}^2+R\mbox{e}^{2(\eta-U)}(\rho+J).$$ 
From (\ref{matt1}) and (\ref{matt3}) we have 
$\rho\geq |J|$, and since $R>0$ 
it follows that both $h_1$ and $h_{2}$ are non-negative. 
Solving equation (\ref{Rxi}) gives for any $\theta_{0}\in [0,2\pi]$ 
(suppressing the $t$-dependence) 
\begin{equation} 
\displaystyle 
R_{\xi}(\theta)=\mbox{e}^{\int_{\theta_{0}}^{\theta}\eta_{\xi}(\sigma)
d\sigma}R_{\xi}(\theta_{0})-\int_{\theta_{0}}^{\theta}
\mbox{e}^{\int_{\tilde{\theta}}^{\theta}\eta_{\xi}(\sigma)d\sigma}
h_1(\tilde{\theta})d\tilde{\theta}.\label{sepxi}
\end{equation} 
Since $R$ is $C^{\infty}$ on $S^1$ it can be identified with a periodic 
function on the real line. If now $R_{\xi}(\theta_{0})=0$ for any $\theta_{0}$ 
then $R_{\xi}(2\pi +\theta_{0})=0$, but from (\ref{sepxi}) this is 
only possible 
if $h_1$ vanishes identically. However, in the non-vacuum case 
(recall $f_0\not= 0$) $h_{\xi}(t,\cdot)$ is strictly positive 
on some open set of $[0,2\pi]$. 
Therefore $R_{\xi}$ is nonzero and has a definite sign. The same arguments 
apply to $R_{\lambda}$, and it follows that 
$$g^{\mu\nu}\partial_{\mu}R\partial_{\nu}R=\mbox{e}^{-2(\eta-U)}
R_{\xi}R_{\lambda}$$ 
is strictly positive or strictly negative. The former possibility is ruled out 
since $\partial_{\theta}R=0$ at some point on $S^1$. 
Thus $\nabla R$ is timelike. This means that $\partial_{t}R$ is nonzero 
everywhere. Our choice of time corresponds to contracting $T^2$ orbits 
so that $\partial_{t}R>0$. 

Next we show that $\partial_{t}R$ and $|\partial_{\theta}R|$ are bounded 
into the past. The evolution equation (\ref{evol2con}) can be written 
\begin{equation}
\partial_{\lambda}R_{\xi}=R\mbox{e}^{2(\eta-U)}(\rho-P_1),\label{xilam}
\end{equation}
or equivalently, 
\begin{equation}
\partial_{\xi}R_{\lambda}=R\mbox{e}^{2(\eta-U)}(\rho-P_1).\label{lamxi}
\end{equation} 
The right hand side is positive since $\rho\geq P_1$, 
see (\ref{matt1}) and (\ref{matt2}), and from (\ref{xilam}) 
it follows that if we start at any point $(t_0,\theta_{0})$ on the initial 
surface we obtain 
\begin{equation} 
R_{\xi}(\theta_{0}+s,t_0-s)\leq R_{\xi}(t_0,\theta_{0}),\label{mindre1} 
\end{equation}
and similarly from (\ref{lamxi}) 
\begin{equation}
R_{\lambda}(\theta_{0}-s,t_0-s)\leq R_{\lambda}(t_0,\theta_{0}).
\label{mindre2}
\end{equation} 
From these relations we get for any $t\in(\tilde{t},t_0)$ and 
any $\theta\in S^1,$ 
\begin{eqnarray}
&\displaystyle R_{\xi}(t,\theta)\leq\sup_{\theta\in S^1}R_{\xi}(t_0,\theta),&\\ 
&\displaystyle R_{\lambda}(t,\theta)\leq\sup_{\theta\in S^1}R_{\lambda}(t_0,\theta).& 
\end{eqnarray} 
This yields 
\begin{equation}
R_{t}(t,\theta)\leq\sup_{\theta\in S^1}(R_{\xi}+R_{\lambda})(t_0,\theta), 
\end{equation}
and since $\nabla R$ is timelike everywhere we have $|R_t|>|R_{\theta}|$ and 
we find that both $R_t$ and $|R_{\theta}|$ are bounded into the past, 
so $R$ is uniformly $C^1$ bounded to the past of the initial surface. 

\textit{Step 2. }(Bounds on $U, A$ and $\eta$ and their first derivatives.)\\ 
The bounds on $U_{t},A_{t},U_{\theta}$ and $A_{\theta}$ to the past 
of the initial surface 
are obtained by a light-cone estimate, which in this case, with one spatial 
dimension, is an application of the Gronwall method on two independent 
null paths. Then, by combining these results, 
one obtains the desired estimate. 

The functions involved in the light-cone argument are quadratic functions 
in the first order derivatives of $U$ and $A,$ defined by 
\begin{eqnarray} 
G&=&\frac{1}{2}R(U_{t}^2+U_{\theta}^2)+\frac{\mbox{e}^{4U}}{8R}
(A_{t}^2+A_{\theta}^2),\label{G}\\ 
H&=&RU_{t}U_{\theta}+\frac{\mbox{e}^{4U}}{4R}A_tA_{\theta}.\label{H} 
\end{eqnarray} 
We will see below that if we let the vector fields along the null paths 
act on $G+H$ and $G-H$ we obtain 
equations appropriate for applying a 
Gronwall argument. First, however, we motivate the definition 
of $G$ and $H$ by showing that they 
are components of an ``enegy-momentum tensor'' of a wave map. 
So let us follow [BCIM] and give 
a brief discussion of such a map in this context. 
Consider a base Lorentzian manifold $(\mathcal{N},\nu)$ with 
the two dimensional manifold $\mathcal{N}$ corrseponding to the past 
conformal coordinate development of $(h,k,f_0)$ with the metric 
$$\nu:=-dt^2+d\theta^2,$$ and consider a family of target manifolds 
$(\mathbb{R}^2,d_{(t,\theta)})$ with 
$$d_{(t,\theta)}:=R(t,\theta)dU^2+\frac{\mbox{e}^{4U}}{4R(t,\theta)}dA^2.$$ 
Define the map 
\begin{eqnarray} 
\displaystyle\Phi&:&\mathcal{N}\rightarrow\mathbf{R^2}, 
\nonumber\\ 
\displaystyle & &(t,\theta)\mapsto \Phi(t,\theta)=(U(t,\theta),A(t,\theta)), 
\end{eqnarray} 
with $d_{(t,\theta)}$ providing an inner product on their tangents, e.g. 
$$<\Phi_{t},\Phi_{\theta}>=d_{ab}\Phi_{t}^a\Phi_{\theta}^b=RU_{t}U_{\theta}
+\frac{\mbox{e}^{4U}}{4R}A_{t}A_{\theta}.$$ 
There is a covariant derivative $D$ compatible with $\nu$ and 
semi-compatible with $d_{(t,\theta)}$. Using Greek indices for the base and 
Latin indices for the target the action $D$ can be expressed as 
\begin{equation} 
D_{\mu}\Phi_{\lambda}^a=\partial_{\mu}\Phi_{\lambda}^a
+\Gamma_{bc}^a\Phi_{\lambda}^b\Phi_{\mu}^c-\Gamma_{\lambda\mu}^{\kappa}
\Phi_{\kappa}^a, 
\end{equation} 
with 
$$\Gamma_{UU}^U=0,
\;\;\;\Gamma_{UA}^U=0,\;\;\;\Gamma_{AA}^U=-\frac{\mbox{e}^{4U}}{2R^2},$$ 
$$\Gamma_{AA}^A=0,\;\;\;\Gamma_{AU}^A=2,\;\;\;\Gamma_{UU}^A=0,$$ 
and the base Christoffel coefficients $\Gamma_{\lambda\mu}^{\kappa}=0$ 
(in the expanding direction these will be nonzero). 
As pointed out above, $D$ is compatible with the flat metric $\nu$ 
but not with $d_{(t,\theta)}$, 
we have 
$$D_{\lambda}d_{ab}=R_{\lambda}(\delta_{a}^U\delta_{b}^U-\frac{\mbox{e}^{4U}}{4R^2}
\delta_{a}^A\delta_{b}^A).$$ 
By defining the wave operator $\Box:=\nu^{\lambda\mu}
D_{\lambda}D_{\mu}$ the evolution equations for $U$ and $A$ take the form 
\begin{eqnarray} 
\displaystyle\Box U&=&\frac{U_{t}R_{t}}{R}-\frac{U_{\theta}R_{\theta}}{R}
+\frac{1}{2}\mbox{e}^{2(\eta-U)}(\rho-P_1+P_2-P_3)
\nonumber\\ 
\displaystyle\Box A&=&-\frac{A_{t}R_{t}}{R}+\frac{A_{\theta}R_{\theta}}{R}
+2R\mbox{e}^{2(\eta-2U)}S_{23}.\nonumber
\end{eqnarray} 
Now we define an ``energy-momentum tensor'' for the maps $\Phi,$ 
\begin{eqnarray} 
\displaystyle T_{\lambda\mu}&=&<\Phi_{\lambda},\Phi_{\mu}>-\frac{1}{2}
\nu_{\lambda\mu}\nu^{\delta\gamma}<\Phi_{\delta},\Phi_{\gamma}>
\nonumber\\ 
\displaystyle &=&RU_{\lambda}U_{\mu}+\frac{\mbox{e}^{4U}}{4R}A_{\lambda}A_{\mu}
+\frac{1}{2}\nu_{\lambda\mu}[R(U_{t}^2-U_{\theta}^2)
+\frac{\mbox{e}^{4U}}{4R}(A_{t}^2-A_{\theta}^2)].\nonumber 
\end{eqnarray} 
We find that $G=T_{tt}$ and $H=T_{t\theta}$. 

Let us now derive bounds on $U$ and $A$ and their first order derivatives. 
By using the evolution equation (\ref{evol1con}) and (\ref{evol4con}) 
we find 
\begin{eqnarray}
\displaystyle\partial_{\lambda}(G+H)&=&\frac{-1}{2\sqrt{2}}R_{\xi}
\left(U_{t}^2-U_{\theta}^2+\frac{\mbox{e}^{4U}}{4R^2}
(-A_{t}^2+A_{\theta}^2)\right)
\nonumber\\
\displaystyle & &+\frac{R}{2}U_{\xi}
\mbox{e}^{2(\eta-U)}(\rho-P_1+P_2-P_3)
\nonumber\\
\displaystyle & &+\frac{\mbox{e}^{2U}}{2R}A_{\xi}
R\mbox{e}^{2(\eta-U)}S_{23},\nonumber 
\end{eqnarray}
and 
\begin{eqnarray} 
\displaystyle\partial_{\xi}(G-H)&=&\frac{-1}{2\sqrt{2}}R_{\lambda}
\left(U_{t}^2-U_{\theta}^2+\frac{\mbox{e}^{4U}}{4R^2}
(-A_{t}^2+A_{\theta}^2)\right)
\nonumber\\
\displaystyle & &+\frac{R}{2}U_{\lambda} 
\mbox{e}^{2(\eta-U)}(\rho-P_1+P_2-P_3)
\nonumber\\ 
\displaystyle & &+\frac{\mbox{e}^{2U}}{2R}A_{\lambda}
R\mbox{e}^{2(\eta-U)}S_{23}.\nonumber 
\end{eqnarray} 
Now, integrating these equations along null paths starting at $(t_1,\theta)$ 
and ending at the initial $t_0-$surface, and adding the results we obtain 
\begin{eqnarray} 
&\displaystyle G(t_1,\theta)=\frac{1}{2}[G+H](t_0,\theta-(t_0-t_1))
+\frac{1}{2}[G+H](t_0,
\theta+(t_0-t_1))&\nonumber\\ 
&\displaystyle -\frac{1}{2}\int_{t_1}^{t_0} 
K_{1}(s,\theta-(s-t_1))+K_{2}(s,\theta+(s-t_1))\,ds&\nonumber\\ 
&\displaystyle -\frac{1}{2}\int_{t_1}^{t_0}
[U_{\xi}T](s,\theta-(s-t_1))+ 
[U_{\lambda}T](s,\theta+(s-t_1))\,ds& 
\nonumber\\ 
&\displaystyle -\frac{1}{2}\int_{t_1}^{t_0}
[\frac{\mbox{e}^{2U}}{2R}A_{\xi}\tilde{T}](s,\theta-(s-t_1))
+[\frac{\mbox{e}^{2U}}{2R}A_{\lambda}\tilde{T}](s,\theta+(s-t_1))\,ds,& 
\label{gvnyc} 
\end{eqnarray} 
where we have introduced the notations 
\begin{eqnarray} 
&\displaystyle K_{1}=\frac{-1}{2\sqrt{2}}
R_{\lambda}\left((U_{t}^2-U_{\theta}^2+\frac{\mbox{e}^{4U}}{R^2}
(-A_{t}^2+A_{\theta}^2)\right),&\\ 
&\displaystyle K_{2}=\frac{-1}{2\sqrt{2}}R_{\xi}\left((U_{t}^2-U_{\theta}^2
+\frac{\mbox{e}^{4U}}{R^2}
(-A_{t}^2+A_{\theta}^2)\right),&\\
&\displaystyle T=\frac{R}{2}\mbox{e}^{2(\eta-U)}(\rho-P_1+P_2-P_3)&\\ 
&\displaystyle \tilde{T}=R\mbox{e}^{2(\eta-U)}S_{23}.&
\end{eqnarray} 
Let us first consider the matter terms. Note that for 
any $t\in(\tilde{t},t_0)$, the evolution equations 
(\ref{xilam}) and (\ref{lamxi}) give 
\begin{equation}
\displaystyle R_{\xi}(t_0,\theta+(t_0-t))-R_{\xi}(t,\theta)=
\sqrt{2}\int_{t}^{t_0}[R\mbox{e}^{2(\eta-U)}(\rho-P_1)](s,\theta+(s-t))ds, 
\label{mati1} 
\end{equation} 
and 
\begin{equation}
\displaystyle R_{\lambda}(t_0,\theta-(t_0-t))-R_{\xi}(t,\theta)=
\sqrt{2}\int_{t}^{t_0}[R\mbox{e}^{2(\eta-U)}(\rho-P_1)]
(s,\theta-(s-t))ds.\label{mati2} 
\end{equation} 
Hence, since $R$ is uniformly $C^1$ bounded to the past of the initial surface 
it follows that the right hand sides are uniformly bounded. 
From (\ref{matt1})-(\ref{matt2}) we have $$\rho\geq P_1+P_2+P_3,$$ and 
thus $$0\leq (\rho-P_1+P_2-P_3)\leq 2(\rho-P_1),$$ and 
from (\ref{matt4}) and the elementary inequatlity 
$2ab\leq a^2+b^2,\, a,b\in\mathbb{R},$ 
we have $$2|S_{23}|\leq P_2+P_3\leq\rho-P_1.$$ 
In view of (\ref{mati1}) and (\ref{mati2}) we therefore have 
that both 
\begin{equation} 
\displaystyle\int_{t}^{t_0}T(s,\theta\pm (s-t))ds, 
\end{equation} 
and 
\begin{equation} 
\displaystyle\int_{t}^{t_0}|\tilde{T}(s,\theta\pm (s-t))|ds, 
\end{equation} 
are uniformly bounded on $(\tilde{t},t_0]\times S^1$, say by $C(t)$. 
Now, by using the inequality $2ab\leq a^2+b^2$ again, we get 
$$|U_{\xi}|\leq\left(\frac{2G}{R}\right)^{1/2},$$ and 
$$\frac{\mbox{e}^{2U}}{2R}|A_{\xi}|\leq\left(\frac{2G}{R}\right)^{1/2}.$$ 
The same estimates also hold for $U_{\lambda}$ and $A_{\lambda}$. 
Since $R_{\xi}$ and $R_{\lambda}$ are uniformly bounded it clearly follows 
that $$|K_1|\leq\frac{CG}{R},\;\;|K_2|\leq\frac{CG}{R},$$ 
for some constant $C$. 
So the identity (\ref{gvnyc}) now implies, as long as $R$ stays bounded 
away from zero, that 
\begin{eqnarray}
&\displaystyle\sup_{\theta}G(t_1,\theta)\leq\sup_{\theta}
2G(t_0,\theta)+\sup_{\theta}2G(t_0,\theta)&\\
&\displaystyle+C\sup_{[t_1,t_0]\times S^1}(G(\cdot,\cdot))^{1/2} 
+\int_{t_1}^{t_0}C\sup_{\theta}G(s,\cdot)ds.& 
\end{eqnarray} 
On a sub time interval $[\sigma_1,\sigma_0]\subset [t_1,t_0]$ on which 
$\sup_{S^1}G(s,\cdot)$ is increasing as $s\rightarrow\sigma_1$, which is 
the case of interest for us since we wish to obtain a uniform upper bound 
on $G$, 
we have $$\sup_{[\sigma_1,\sigma_0]\times S^1}(G(\cdot,\cdot))^{1/2}\leq
\sup_{S^1}(G(\sigma_1,\cdot))^{1/2}.$$ Therefore, by dividing both sides by 
$\sup_{S^1}(G(\sigma_1,\cdot))^{1/2}$ we obtain a Gronwall inequality on 
the time interval $[\sigma_1,\sigma_0]$ for $\sup_{S^1}G^{1/2}$ leading 
to a upper bound on $\sup_{S^1}G(\sigma_1,\cdot)$ as long as $R$ stays bounded away from zero. 
By repeating this argument we conclude that $\mbox{sup}_{S^1}G$ is 
uniformly bounded on $(\tilde{t},t_0]$, leading to bounds on $U$ and its 
first order derivatives, and thus also 
on $A$ and its first 
order derivatives, as long as $R$ stays bounded away from zero. 
The bounds on $|\eta|, |\eta_{t}|$ 
and $|\eta_{\theta}|$ are obtained in a similar way since the evolution 
equation (\ref{evol3con}) can be written 
\begin{equation} 
\displaystyle\partial_{\lambda}\eta_{\xi}=U_{\theta}^2-U_{t}^2
+\frac{\mbox{e}^{4U}}{4R^2}(A_{t}^2-A_{\theta}^2)-\mbox{e}^{2(\eta-U)}
(P_{3}+\frac{A^2}{R^2}\mbox{e}^{4U}P_2+\frac{2A}{R}\mbox{e}^{2U}S_{23}), 
\end{equation} 
or equivalently, 
\begin{equation} 
\displaystyle\partial_{\xi}\eta_{\lambda}=U_{\theta}^2-U_{t}^2+\frac{\mbox{e}^{4U}}{4R^2}(A_{t}^2-A_{\theta}^2)-\mbox{e}^{2(\eta-U)}
(P_{3}+\frac{A^2}{R^2}\mbox{e}^{4U}P_2+\frac{2A}{R}\mbox{e}^{2U}S_{23}). 
\end{equation} 
We found above that the integrals along null paths for the matter quantity 
$R\mbox{e}^{2(\eta-U)}(\rho-P_1)$ were bounded to the past of the initial 
surface. Therefore, since $0\leq P_k\leq\rho-P_1,\; k=2,3$ and 
$|S_{23}|\leq\rho-P_1$ we have, as long as $R$ 
stays bounded away from zero, that the integrals along the null paths 
for the matter terms in the rigt hand sides above are bounded as well, 
since $U$ and $A$ are bounded. 
Now, since the first order derivatives of $U$ and $A$ are uniformly bounded 
we immediately obtain that $|\eta_{\xi}|$ and 
$|\eta_{\lambda}|$ are bounded by integrating the equations for $\eta$ along 
null paths. Since $\eta_{t}=\frac{1}{\sqrt{2}}(\eta_{\xi}+\eta_{\lambda})$ 
and $\eta_{t}=\frac{1}{\sqrt{2}}(\eta_{\xi}-\eta_{\lambda})$ we find that 
$\eta$ is uniformly $C^1$ bounded to the past of the initial surface 
as long as $R$ stays bounded away from zero. 

\textit{Step 3. }(Bound on the support of the momentum.)\\ 
Note that a solution $f$ to the Vlasov equation is given by 
\begin{equation} 
f(t,\theta,v)=f_{0}(\Theta(0,t,\theta,v),V(0,t,\theta,v)),\label{solut} 
\end{equation} 
where $\Theta$ and $V$ are solutions to the characteristic system 
\begin{eqnarray} 
\displaystyle\frac{d\Theta}{ds}&=&\frac{V^1}{V^0}, 
\nonumber\\ 
\displaystyle\frac{dV^1}{ds}&=&-
(\eta_{\theta}-U_{\theta})V^{0}-(\eta_{t}-U_{t})V^{1}+U_{\theta}
\frac{(V^{2})^{2}}{V^{0}}
\nonumber\\
\displaystyle & &-(U_{\theta}-\frac{R_{\theta}}{R})\frac{(V^3)^2}{V^0}
+\frac{A_{\theta}}{R}
\mbox{e}^{2U}\frac{V^2V^3}{V^0}, 
\nonumber\\ 
\displaystyle\frac{dV^2}{ds}&=&-U_{t}V^{2}-U_{\theta}
\frac{V^{1}V^{2}}{V^{0}}, 
\nonumber\\ 
\displaystyle\frac{dV^3}{ds}&=&-(\frac{R_{t}}{R}-U_{t})V^{3}
+(U_{\theta}-\frac{R_{\theta}}{R})\frac{V^{1}V^{3}}{V^{0}}
\nonumber\\ 
\displaystyle & &-\frac{\mbox{e}^{2U}
}{R}(A_t+A_{\theta}\frac{V^1}{V^0})V^2, 
\nonumber 
\end{eqnarray} 
and $\Theta(s,t,x,v),\;V(s,t,x,v)$ is the solution that goes 
through the point $(\theta,v)$ at time $t$. 
Let us recall the definition of
$$Q(t):=\mbox{sup}\{|v|:\exists (s,\theta)\in [t,t_0]\times
S^1\mbox{such that} f(s,\theta,v)\not= 0\}.$$ If $Q(t)$ can be controlled we
obtain immediately from (\ref{matt1})-(\ref{matt3}) bounds
on $\rho, J,S_{23}$ and $P_k,\, k=1,2,3$, since $\|f\|_{\infty}\leq
\|f_0\|_{\infty}$ from (\ref{solut}). 
Now, all of the field components and their first derivatives are known to be
bounded on $(\tilde{t},t_0]$, as long as $R$ stays bounded away from zero. 
Also, the distribution function has compact support on the initial surface 
and therefore $|V^k(t_{0})|<C.$ 
So by observing that $|V^k|<V^0$, $k=1,2,3,$ a simple Gronwall argument 
applied to the characteristic system gives 
uniform bounds on $|V^{k}(t)|,$ $t\in(\tilde{t},t_0]$, 
and it follows that $Q(t)$ is uniformly bounded on 
$(\tilde{t},t_0]$. 

\textbf{Remark. }By a Killing vector argument, bounds on $|V^2|$ and $|V^3|$ 
can be derived if merely $|U|$ and $|A|$ are bounded and $R>\epsilon>0$. 
Such an argument will be 
used in the expanding direction.\\ 


\textit{Step 4. }(Bounds on the second order derivatives of the field 
components and on the first order derivatives of $f$.) 
From the Einstein-matter constraint equations in conformal coordinates 
we can express $R_{t\theta}$ and $R_{\theta\theta}$ in terms of 
uniformly bounded 
quantities, as long as $R$ stays bounded away from zero. Therefore these 
functions are uniformly bounded and 
equation (\ref{evol2con}) then implies that $R_{tt}$ is uniformly 
bounded as well. 

In the vacuum case one can take the derivative of the evolution equations and 
repeat the argument in step 2 to obtain bounds on second order derivatives 
of $U$ and $A$. Here we need another argument. 
First we write the evolution equations for $U$ 
and $A$ in the forms 
\begin{eqnarray}
\displaystyle U_{tt}-U_{\theta\theta}&=&\frac{(R_{\theta}-R_{t})}{2R}(U_{\theta}+U_{t})
-\frac{(R_{\theta}+R_{t})}{2R}(U_{t}-U_{\theta}) 
\nonumber\\ 
\displaystyle & &+\frac{\mbox{e}^{4U}}{2R^2}(A_t-A_{\theta})(A_t+A_{\theta}) 
+\frac{1}{2}\mbox{e}^{2(\eta-U)}\kappa, 
\end{eqnarray} 
and 
\begin{eqnarray} 
\displaystyle A_{tt}-A_{\theta\theta}&=&\frac{(R_{t}-R_{\theta})}{2R}
(A_{\theta}+A_{t})
+\frac{(R_{\theta}+R_{t})}{2R}(A_{t}-A_{\theta})
\nonumber\\ 
\displaystyle & &-2(A_{t}-A_{\theta})(U_{\theta}+U_{t})
-2(A_{\theta}+A_{t})(U_{t}-U_{\theta})
\nonumber\\ 
\displaystyle & &+2R\mbox{e}^{2(\eta-2U)}S_{23}, 
\end{eqnarray} 
where $\kappa$ denotes $\rho-P_1+P_2-P_3$. 
Taking the $\theta$-derivative of these equations gives 
\begin{eqnarray} 
\partial_{\lambda}\partial_{\xi}U_{\theta}&=&L+\frac{R_{\lambda}}{2R}
\partial_{\xi}
U_{\theta}+ 
\frac{R_{\xi}}{2R}\partial_{\lambda}U_{\theta}
\nonumber\\ 
\displaystyle & &+\frac{\mbox{e}^{4U}}{2R^2}
(A_{\lambda}\partial_{\xi}A_{\theta}+A_{\xi}\partial_{\lambda}A_{\theta})+ 
\frac{1}{4}\mbox{e}^{2(\eta-U)}\kappa_{\theta},\label{dthe1} 
\end{eqnarray} 
and 
\begin{eqnarray} 
\partial_{\lambda}\partial_{\xi}A_{\theta}&=&L+\frac{R_{\lambda}}{2R}
\partial_{\xi}
A_{\theta}-\frac{R_{\xi}}{2R}\partial_{\lambda}A_{\theta}+ 
2U_{\xi}\partial_{\lambda}A_{\theta}+2A_{\lambda}\partial_{\xi}U_{\theta}
\nonumber\\ 
\displaystyle & &+2U_{\lambda}\partial_{\xi}A_{\theta}+2A_{\xi}\partial_{\lambda}U_{\theta}+2R\mbox{e}^{2(\eta-2U)}(S_{23})_{\theta}, 
\end{eqnarray} 
Here, $L$ contains only $\kappa$ and $S_{23}$, 
first order derivatives of $U, A$ and $\eta$, and first and second order 
derivatives of $R$, which all are known to be bounded. 
These equations can of course also be written in a form where 
the left hand sides read $\partial_{\xi}\partial_{\lambda}U_{\theta}$ 
and $\partial_{\xi}\partial_{\lambda}A_{\theta},$ respectively. 
By integrating these equations along null paths to the past of the initial 
surface, we get from a Gronwall argument a bound 
on $$\sup_{\theta\in S^1}(|\partial_{\xi}U_{\theta}|
+|\partial_{\lambda}U_{\theta}|+|\partial_{\xi}A_{\theta}|
+|\partial_{\lambda}A_{\theta}|),$$ as long 
as $R$ is bounded away from zero, under the hypothesis that the integral 
of the differentiated matter terms $\kappa_{\theta}$ and $(S_{23})_{\theta}$ 
can be controlled. 
In order to bound these integrals we make use of a device introduced 
by Glassey and Strauss [GS] for treating the Vlasov-Maxwell equation. 
It is sufficient to show how one of the differentiated matter terms can 
be boundeded since the arguments are similar in all cases. Let us consider 
the integral appearing by integrating (\ref{dthe1}) along the null path 
defined by $\partial_{\lambda}$ 
which involves $\rho_{\theta}$, 
\begin{equation} 
\frac{1}{4}\int_{t}^{t_0}\int_{\mathbb{R}^3}[\mbox{e}^{2(\eta-U)}v^{0}
\partial_{\theta}f] 
(s,\theta-(s-t),v)dvds, 
\end{equation} 
where $t\in (\tilde{t},t_0]$. Next, define 
$$W=\sqrt{2}\partial_{\lambda}=\partial_{t}-\partial_{\theta},\;\;\;\, 
S=\partial_{t}+\frac{v^1}{v^0}\partial_{\theta}.$$ 
Hence, $\partial_{\theta}$ and $\partial_{t}$ can be expressed in 
terms of $W$ and $S$ by 
\begin{eqnarray} 
\partial_{\theta}&=&\frac{v^0}{v^0+v^1}(S-W),\label{inver}\\ 
\partial_{t}&=&\frac{v^0}{v^0+v^1}(S+\frac{v^1}{v^0}W). 
\end{eqnarray} 
Now, $$[Wf](s,\theta-(s-t),v)=\partial_{s}[f(s,\theta-(s-t),v)],$$ and from 
the Vlasov equation we get 
$$[Sf](s,\theta-(s-t),v)=[-K\cdot\nabla_{v}f](s,\theta-(s-t),v),$$ 
where it is clear which terms have been denoted by $K=(K_1,K_2,K_3)$. 
By using (\ref{inver}) we can now evaluate the integral above by 
integrating by parts (in $s$ for the $W$-term and in $v$ for the $S$-term), so 
that the remaining terms only involve bounded quantities. Note in particular 
the $v$-integrals are easily controlled in view of the uniform bound 
on $Q(t)$. 
Thus, the integrals of the differentiated matter terms can 
be controlled and the Gronwall argument referred to above goes through. 
So we obtain uniform bounds 
on $|\partial_{\xi}U_{\theta}|, |\partial_{\lambda}U_{\theta}|, 
|\partial_{\xi}A_{\theta}|$, and $|\partial_{\lambda}A_{\theta}|,$ and 
therefore also on $|U_{\theta\theta}|, |U_{t\theta}|, |A_{\theta\theta}|$ 
and $|A_{t\theta}|,$ as long as $R$ is 
bounded away from zero. The evolution 
equations (\ref{evol1con}) and (\ref{evol4con}) then give uniform bounds 
on $|U_{tt}|$ and $|A_{tt}|$. 
By differentiating equation (\ref{evol3con}), it is now straightforward to 
obtain bounds on the second order derivatives of $\eta$, using similar 
arguments to those already discussed here, in particular the integrals 
involving matter quantities can be treated as above. Bounds 
on the first order derivatives of the distribution function $f$ may now be 
obtained from the known bounds on the field components from the formula 
\begin{equation}
f(t,\theta,v)=f_{0}(\Theta(0,t,\theta,v),V(0,t,\theta,v)), 
\end{equation}\\ 
since $f_0$ is smooth and since $\partial\Theta$ and 
$\partial V$ (here $\partial$ denotes $\partial_{t}, 
\partial_{\theta}$ or $\partial_v$) can be controlled by a Gronwall argument 
in view of the characteristic system. 

\textit{Step 5. }(Bounds on higher order derivatives and completion 
of the proof.) 
It is clear that the method described above can be continued for 
obtaining bounds on higher derivatives as well. Hence, we have uniform bounds 
on the functions $R,U,A,\eta$ and $f$ and all their derivatives 
on the interval $(\tilde{t},t_0]$ if $R>\epsilon>0$. 
This implies that the solution 
extends to $t\rightarrow -\infty$ as long as $R$ stays bounded away from zero. 
In view of the discussion after the statement of Theorem 1, this 
completes the proof of Theorem 1 in the contracting direction. 
\begin{flushright}
$\Box$ 
\end{flushright} 
\section{Analysis in the expanding direction} 
To begin the analysis in the expanding direction (increasing $R$) in 
areal coordinates we need to start with data 
on a $R=$constant Cauchy surface (recall that in areal coordinates $R=t$). 
That this can be done follows from the geometrical arguments in [BCIM] 
(cf. the discussion following the statement of Theorem 1).
There it is shown that if Gowdy symmetric (or more generally $T^2$ symmetric) 
data is given on $T^3$, and if $R_0$ is the past limit of $R$ along past 
inextendible paths in $D^{-}_{conf}$ and if $R_1:=\inf_{T^3}R$, 
then for every $d\in(R_0,R_1)$, the $R=d$ level set $\Sigma_{d}$ is a Cauchy 
surface, and these $\Sigma_{d}$ foliate the region $D^{-}_{conf}
\cap I^{-}(\Sigma_{R_1})$. Here $I^{-}(S)$ is the chronological past of $S$ 
(see [HE]). The surfaces $\Sigma_{d}$ lie to the past of the initial surface. 
Let us pick one of them, say $\Sigma_{d_2}$. 
The spacetime $D^{-}(h,k,f_0)$ induces 
initial data for the areal component fields $(U,A,\eta,\alpha)$ 
and the distribution function $f$ on $\Sigma_{t_{2}=d_{2}}$. By combining 
the local existence proof in harmonic coordinates [CB], and the arguments 
in [Cl] which show that the spacetime admits areal coordinates, we obtain 
local existence for the initial value problem in these coordinates. 
Now, in order to extend local existence to global existence in these 
coordinates, it is again sufficient to obtain uniform bounds 
on the field components and the distribution function and all 
their derivatives on a finite time interval $[t_2,t_3)$ on 
which the local solution exists. 

\textit{Step 1. }(Bounds on $\alpha, U, A$ and $\tilde{\eta}$.)\\ 
In this step we first show an ``energy'' monotonicity lemma and then 
we show how this result leads to bounds on $\tilde{\eta}:=\eta+\ln{\alpha}/2$ 
and on $U$ and $A$. 
Let $E(t)$ be defined by 
$$E(t)=\int_{S^1}[\alpha^{-\frac{1}{2}}U_{t}^2+\sqrt{\alpha}
U_{\theta}^2+\frac{\mbox{e}^{4U}}{4t^2}(\alpha^{-\frac{1}{2}}A_{t}^2+\sqrt{\alpha}
A_{\theta}^2)+\sqrt{\alpha}\mbox{e}^{2(\eta-U)}\rho] d\theta.$$ 
\begin{lemma} 
$E(t)$ is a monotonically decreasing function in $t$, and satisfies
\begin{equation} 
\frac{d}{dt}E(t)=-\frac{2}{t}\int_{S^1}[\alpha^{-1/2}U_{t}^2 
+\frac{\mbox{e}^{4U}}{4t^2}\sqrt{\alpha}A_{\theta}^2+\frac{\sqrt{\alpha}}{2} 
\mbox{e}^{2(\eta-U)}(\rho+P_3)]d\theta\leq 0.\label{energ} 
\end{equation} 
\end{lemma} 
\textit{Proof. }This is a straightforward computation, using the evolution 
equation for $U$ and $A$, the constraint equations and the Vlasov equation. 
First, using the evolution equations for $U$ and $A$ we 
obtain after a short computation 
\begin{eqnarray} 
&\displaystyle\frac{d}{dt}\int_{S^1}[\alpha^{-\frac{1}{2}}U_{t}^2+\sqrt{\alpha}U_{\theta}^2+\frac{\mbox{e}^{4U}}{4t^2}(\alpha^{-1/2}A_{t}^2+\sqrt{\alpha}A_{\theta}^2)] d\theta& 
\nonumber\\ 
&\displaystyle =\int_{S^1}\frac{\alpha_{t}}{2\alpha^{3/2}}\left(
U_{t}^2+\alpha U_{\theta}^2+\frac{\mbox{e}^{4U}}{4t^2}
(A_{t}^2+\alpha A_{\theta}^2)\right)-\frac{2}{t}\left(
\frac{U_{t}^2}{\sqrt{\alpha}}
+\frac{\mbox{e}^{4U}}{4t^2}\sqrt{\alpha}A_{\theta}^2\right)d\theta& 
\nonumber\\ 
&\displaystyle +\int_{S^1}\sqrt{\alpha}U_{t}\mbox{e}^{2(\eta-U)}
(\rho-P_1+P_2-P_3)+\frac{\mbox{e}^{2\eta}}{t}\sqrt{\alpha}
A_{t}S_{23}\,d\theta.&\label{part1} 
\end{eqnarray} 
Next we have 
\begin{eqnarray} 
&\displaystyle\frac{d}{dt}\int_{S^1}\sqrt{\alpha}\mbox{e}^{2(\eta-U)}
\rho d\theta=\int_{S^1}\frac{\alpha_{t}}{2\sqrt{\alpha}}\mbox{e}^{2(\eta-U)}
\rho d\theta&\\ 
&\displaystyle +\int_{S^1}2\sqrt{\alpha}(\eta_{t}-U_{t})\mbox{e}^{2(\eta-U)}
\rho d\theta+ 
\int_{S^1}\sqrt{\alpha}\mbox{e}^{2(\eta-U)}\rho_{t}d\theta.&\label{part2} 
\end{eqnarray} 
Now we use the Vlasov equation and the definition of $\rho$ to 
write the last integral as 
\begin{eqnarray} 
&\displaystyle\int_{S^1}\sqrt{\alpha}\mbox{e}^{2(\eta-U)}\rho_{t}d\theta= 
\int_{S^1}\int_{\mathbb{R}^3}\sqrt{\alpha}\mbox{e}^{2(\eta-U)}
[-\sqrt{\alpha}v^{1}\partial_{\theta}f]dvd\theta&\nonumber\\ 
&\displaystyle +\int_{S^1}\int_{\mathbb{R}^3}\sqrt{\alpha}\mbox{e}^{2(\eta-U)}
\left[(\eta_{\theta}-U_{\theta}
+\frac{\alpha_{\theta}}{2\alpha})\sqrt{\alpha}(v^0)^2 
+(\eta_{t}-U_{t})v^1v^0\right.& 
\nonumber\\ 
&\displaystyle \left.
+\sqrt{\alpha}U_{\theta}[(v^3)^2-(v^2)^2]-\frac{\mbox{e}^{2U}}{t}
\sqrt{\alpha}A_{\theta}v^2v^3\right] 
\frac{\partial f}{\partial v^1} dvd\theta&\nonumber\\ 
&\displaystyle +\int_{S^1}\int_{\mathbb{R}^3}\sqrt{\alpha}\mbox{e}^{2(\eta-U)}
[U_{t}v^2v^0+\sqrt{\alpha}U_{\theta}v^1v^2]\frac{\partial f}{\partial v^2}dvd\theta&
\nonumber\\ 
&\displaystyle +\int_{S^1}\int_{\mathbb{R}^3}\sqrt{\alpha}\mbox{e}^{2(\eta-U)}
\left[(\frac{1}{t}-U_{t})v^3v^0-\sqrt{\alpha}U_{\theta}v^1v^3)\right.& 
\nonumber\\ 
&\displaystyle \left.+\frac{\mbox{e}^{2U}}{t}
(v^0A_{t}+\sqrt{\alpha}A_{\theta}v^1)v^2\right]
\frac{\partial f}{\partial v^3}dvd\theta. 
\end{eqnarray} 
Integrating by parts (using the periodicity in $\theta$ and the compact 
support in $v$) and rearranging the terms we get 
\begin{eqnarray} 
&\displaystyle\int_{S^1}\sqrt{\alpha}\mbox{e}^{2(\eta-U)}\rho_{t}d\theta= 
-\int_{S^1}\int_{\mathbb{R}^3}\sqrt{\alpha}\mbox{e}^{2(\eta-U)}
\left[(v^0+\frac{(v^1)^2}{v^0})\eta_{t}\right.& 
\nonumber\\ 
&\displaystyle\left. -(v^0+\frac{(v^1)^2}{v^0}-\frac{(v^2)^2}{v^0}
+\frac{(v^3)^2}{v^0})U_{t} 
+\frac{\mbox{e}^{2U}}{t}A_{t}\frac{v^2v^3}{v^0}\right]f\, dvd\theta &
\nonumber\\ 
&\displaystyle -\int_{S^1}\int_{\mathbb{R}^3}\frac{\sqrt{\alpha}}{t}\mbox{e}^{2(\eta-U)}
(v^0+\frac{(v^3)^2}{v^0})fdvd\theta.&\label{par25} 
\end{eqnarray} 
By adding (\ref{part1}) and (\ref{part2}), using (\ref{par25}) for the last 
term in (\ref{part2}), we see that the terms involving $U_{t}$ 
and $A_{t}$ vanish and we obtain 
\begin{eqnarray} 
\displaystyle\frac{d}{dt}E(t)&=&\int_{S^1}\frac{\alpha_{t}}{2\alpha^{3/2}}
[U_{t}^2+\alpha U_{\theta}^2+\frac{\mbox{e}^{4U}}{4t^2}(A_{t}^2+\alpha 
A_{\theta}^2)
+\alpha\mbox{e}^{2(\eta-U)}\rho]\, d\theta
\nonumber\\ 
\displaystyle & & 
+\int_{S^1}\sqrt{\alpha}\mbox{e}^{2(\eta-U)}(\rho-P_1)\eta_{t}\, d\theta 
\nonumber\\ 
\displaystyle & &-\int_{S^1}
\frac{2U_{t}^2}{t\sqrt{\alpha}}+\frac{\mbox{e}^{4U}}{2t^3}\sqrt{\alpha}
A_{\theta}^2+\frac{\sqrt{\alpha}}{t}
\mbox{e}^{2(\eta-U)}(\rho+P_3)\,d\theta. 
\end{eqnarray} 
By substituting $\alpha_t$ and $\eta_t$ using the constraint equations 
\begin{equation} 
\frac{\eta_{t}}{t}=U_{t}^{2}+\alpha U_{\theta}^{2}+\frac{\mbox{e}^{4U}}{4t^2}
(A_{t}^2+\alpha A_{\theta}^2)
+\mbox{e}^{2(\eta-U)}\alpha\rho, 
\end{equation} 
and 
\begin{equation} 
\alpha_{t}=2t\alpha^{2}\mbox{e}^{2(\eta-U)}(P_{1}-\rho), 
\end{equation} 
the first two terms cancel, and (\ref{energ}) follows. 
\begin{flushright} 
$\Box$ 
\end{flushright} 
\textbf{Remark. }It is clear from (\ref{energ}) that a Gronwall argument leads 
to a bound on $E(t)$ also on $(0,t_{2}].$ For $T^2$ symmetry and vacuum, which is considered in [BCIM], 
this bound is not available. A natural question is then why the areal 
coordinates in our case has to be discarded in the analysis for the past 
direction. However, the analysis of the characteristic system associated to 
the Vlasov equation in lemma 2 depends on the time direction.\\ 

Let us now define the quantity $\tilde{\eta}$ by 
\begin{equation} 
\tilde{\eta}=\eta+\frac{1}{2}\ln{\alpha}. 
\end{equation} 
From the constraint equation (\ref{constr2}) we get 
\begin{equation} 
\tilde{\eta}_{\theta}=2tU_{t}U_{\theta}+\frac{\mbox{e}^{4U}}{2t}
A_{t}A_{\theta}-t\sqrt{\alpha}\mbox{e}^{2(\eta-U)}J. 
\end{equation} 
Now, from the elementary inequality $|ab|\leq\frac{1}{2c}a^2+2cb^2$, for 
any $a,b,c\in\mathbb{R},\,c>0$, and from the fact that $|J|\leq\rho$, it follows from lemma 1 that  for any $t\in [t_2,t_3)$ 
\begin{equation} 
\int_{S^1}|\tilde{\eta}_{\theta}|d\theta\leq tE(t)\leq tE(t_2). 
\end{equation} 
Hence, for any $\theta_1,\theta_2\in S^1$ and  for any $t\in [t_2,t_3)$ we have 
\begin{equation} 
\displaystyle|\tilde{\eta}(t,\theta_2)-\tilde{\eta}(t,\theta_1)|=|
\int_{\theta_1}^{\theta_2}\tilde{\eta}_{\theta}d\theta| 
\leq\int_{S^1}|\tilde{\eta}_{\theta}|d\theta\leq tE(t_2).\label{varia} 
\end{equation} 
Next, using the constraint equations (\ref{constr1}) and (\ref{constr3}), we find that the time derivative of $\tilde{\eta}$ satisfies 
\begin{equation} 
\tilde{\eta}_{t}=t[U_{t}^2+\alpha U_{\theta}^2+\frac{\mbox{e}^{4U}}{4t^2}(A_{t}^2
+\alpha A_{\theta}^2)+\alpha\mbox{e}^{2(\eta-U)}P_1]
\geq 0. 
\end{equation} 
This relation leads to a control of $\int_{S^1}\tilde{\eta}d\theta$ from 
above, namely 
\begin{eqnarray} 
&\displaystyle\int_{S^1}\tilde{\eta}(t,\theta)d\theta-\int_{S^1}\tilde{\eta}(t_2,\theta)d\theta=\int_{t_2}^t\frac{d}{dt}\left(\int_{S^1}
\tilde{\eta}(s,\theta)d\theta\right) ds& 
\nonumber\\ 
&\displaystyle =\int_{t_2}^t\int_{S^1}\sqrt{\alpha}s
[\frac{U_{t}^2}{\sqrt{\alpha}}
+\sqrt{\alpha}U_{\theta}^2+\frac{\mbox{e}^{4U}}{4s^2}
(\frac{A_{t}^2}{\sqrt{\alpha}}+\sqrt{\alpha}A_{\theta}^2)
+\sqrt{\alpha}\mbox{e}^{2(\eta-U)}P_1]d\theta ds& 
\nonumber\\ 
&\displaystyle\leq\sup_{S^1}\sqrt{\alpha}(t_2,\cdot)
\int_{t_2}^t sE(s)ds
\leq C_1\int_{t_2}^t sE(t_2)ds=C_1E(t_2)(t^2-t_{2}^2)/2.\nonumber\\ 
\label{lowbo} 
\end{eqnarray} 
In the first inequality above we used that $P_1\leq\rho$ and in the second 
that $\alpha$ is a monotonically decreasing function 
in $t$ (see (\ref{constr3})) and $C_1:=\sup_{S^1}\sqrt{\alpha(t_2,\cdot)}$. 
We are 
now in a position to obtain a upper bound  on $\tilde{\eta}$ itself. 
By letting $C_2:=\int_{S^1}\tilde{\eta}(t_2,\theta)d\theta$ we 
get from (\ref{lowbo}) the inequality 
\begin{eqnarray} 
&\displaystyle\frac{1}{2}C_1E(t_2)(t^2-t_{2}^2)+C_2\geq
\int_{S^1}\tilde{\eta}(t,\theta)d\theta&\\ 
&\displaystyle =2\pi\max_{S^1}\tilde{\eta}+\int_{S^1}
(\tilde{\eta}-\max_{S^1}\tilde{\eta})d\theta.& 
\end{eqnarray} 
By applying (\ref{varia}) to the last term we find 
\begin{equation} 
\frac{1}{2}E(t_2)(t^2-t_{2}^2)+C_2\geq 2
\pi\max_{S^1}\tilde{\eta}-2\pi tE(t_2). 
\end{equation} 
Therefore, for some bounded function $C(t)$, we have the upper bound 
\begin{equation} 
\max_{S^1}\tilde{\eta}\leq C(t), 
\end{equation} 
and since $\tilde{\eta}_{t}\geq 0$ 
we conclude that $\tilde{\eta}$ is 
uniformly bounded 
on $S^1\times [t_2,t_3)$.\\ 

\textbf{Remark. }In the analysis below $C(t)$ will always denote a 
uniformly bounded function on $[t_2,t_3)$. Sometimes we introduce other 
functions with the same property only for the purpose of trying to make 
some estimates become more transparent.\\ 

Next we show that the boundedness of $E(t)$, 
together with the constraint equation (\ref{constr3}), lead 
to a bound on $|U|$. 
For any $\theta_1,\theta_2\in S^1$, and $t\in [t_2,t_3)$
we get by H\"{o}lder's inequality 
\begin{eqnarray} 
&\displaystyle|U(t,\theta_2)-U(t,\theta_1)|=\left|\int_{\theta_1}^{\theta_2}
U_{\theta}(t,\theta)d\theta\right|&
\nonumber\\ 
&\displaystyle\leq\left(\int_{\theta_1}^{\theta_2}\alpha^{-1/2}d\theta\right)
^{1/2}\left(\int_{\theta_1}^{\theta_2}\sqrt{\alpha}U_{\theta}^2d\theta\right)
^{1/2}.&\label{Hlr} 
\end{eqnarray} 
The second factor on the right hand side is clearly 
bounded by $(E(t_2))^{1/2}$. For the first factor we use the constraint 
equation (\ref{constr3}). 
This equation can be written as 
\begin{equation} 
\partial_{t}(\alpha^{-1/2})=t\sqrt{\alpha}\mbox{e}^{2(\eta-U)}(\rho-P_1), 
\end{equation} 
so that for $t\in [t_2,t_3)$ 
\begin{equation} 
\alpha^{-1/2}(t,\theta)=\int_{t_2}^{t}s\sqrt{\alpha}
\mbox{e}^{2(\eta-U)}(\rho-P_1)ds+\alpha^{-1/2}(t_2,\theta). 
\end{equation} 
Since $\rho\geq P_1$, the integrand is positive and bounded by the last term 
in the integrand of $E(t)$. 
Letting $C$ denote the supremum 
of $\alpha^{-1/2}(t_2,\cdot)$ over $S^1$ we get 
\begin{eqnarray} 
\int_{\theta_1}^{\theta_2}\alpha^{-1/2}d\theta&\leq&\int_{t_2}^{t}s\int_{S^1} 
\sqrt{\alpha}\mbox{e}^{2(\eta-U)}\rho d\theta ds+2\pi C
\nonumber\\ 
&\leq&E(t_2)(t^2-t_{2}^2)/2+2\pi C. 
\end{eqnarray} 
Hence, for any $\theta_1,\theta_2\in S^1$ we have 
\begin{equation} 
|U(t,\theta_2)-U(t,\theta_1)|\leq C(t). 
\end{equation} 
Next we estimate $\int_{S^1}U(t,\theta)d\theta$. 
Let $C:=\int_{S^1}U(t_2,\theta)d\theta$, 
we get by H\"{o}lder's inequality 
\begin{eqnarray} 
&\displaystyle\left|\int_{S^1}U(t,\theta)d\theta\right|=\left|\int_{t_2}^{t}\int_{S^1}
U_{t}(s,\theta)d\theta ds+C\right|&
\nonumber\\ 
&\displaystyle \leq\int_{t_2}^{t}\int_{S^1}|U_{t}(s,\theta)|d\theta ds+|C|&
\nonumber\\ 
&\displaystyle\leq\int_{t_2}^{t}\left(\int_{S^1}\sqrt{\alpha}d\theta\right)
^{1/2}\left(\int_{S^1}\alpha^{-1/2}U_{t}^2 d\theta\right)^{1/2}ds+|C|.& 
\end{eqnarray} 
The right hand side is easily seen to be bounded since (\ref{constr3}) shows 
that $\sqrt{\alpha}$ is monotonically decreasing and (\ref{energ}) gives 
a bound for the second factor. 
Therefore $$\left|\int_{S^1}U(t,\theta)d\theta\right|\leq C(t),$$ 
for some uniformly bounded function 
$C(t)$. To obtain a uniform bound on $U$ we combine these results. 
Let $U_{+}(t):=\max_{S^1}U(t,\cdot),$ and  $U_{-}(t):=\min_{S^1}U(t,\cdot)$. 
We have 
\begin{equation} 
2\pi U_{\pm}(t)=\int_{S^1}U(t,\theta)d\theta
+\int_{S^1}(U_{\pm}(t)-U(t,\theta))d\theta, 
\end{equation} 
and the right hand side is bounded from below and above so $U$ is 
uniformly bounded on $[t_2,t_3)\times S^1$. These arguments also apply to $A$ 
as well, since the factor $\mbox{e}^{4U}$ is controlled by the uniform bound 
on $U$. 

\textbf{Remark. }In the case studied in [BCIM], i.e. vacuum and $T^2$ 
symmetry, a bound on $\ln{\alpha}$, and thus on $\eta$, is directly 
available. On the other hand, the method used here to bound $U$ 
and $A$ does not apply. This may lead to a difficulty in 
generalizing the result in [BCIM] to matter spacetimes, since the bounds 
on $U$ and $A$ are crucial in order to treat the matter terms 
when the derivatives of $U$ and $A$ are to be bounded. 

\textit{Step 2. }(Bounds on $U_{t},U_{\theta},A_{t},A_{\theta},\eta_{t},
\alpha_{t}$ and $Q(t)$.)\\ 
To bound the derivatives of $U$ we use light-cone estimates in a similar way 
as for the contracting direction. However, the matter terms must be treated 
differently and we need to carry out a careful analysis of the characteristic 
system associated with the Vlasov equation. Let us define 
\begin{eqnarray} 
G&=&\frac{1}{2}(U_{t}^2+\alpha U_{\theta}^2)+\frac{\mbox{e}^{4U}}{8t^2}
(A_{t}^2+\alpha A_{\theta}^2),\label{GE}\\
H&=&\sqrt{\alpha}U_{t}U_{\theta}+\frac{\mbox{e}^{4U}}{4t^2}A_{t}A_{\theta},\label{HE} 
\end{eqnarray} 
and 
\begin{eqnarray} 
\partial_{\chi}&=&\frac{1}{\sqrt{2}}(\partial_{t}+\sqrt{\alpha}
\partial_{\theta})\\ 
\partial_{\zeta}&=&\frac{1}{\sqrt{2}}(\partial_{t}-\sqrt{\alpha}
\partial_{\theta}) 
\end{eqnarray} 
A motivation for the introduction of these quantities is based on similar 
arguments as those given in step 2, section 4. For details we refer to [BCIM]. 

\textbf{Remark. }We use the same notations, $G$ and $H$, 
as in the contracting direction, and below we continue to carry over 
the notations. The analysis in the respective direction is independent 
so there should be no risk of confusion.\\ 

By using the evolution equation (\ref{evol2}), a short computation shows that 
\begin{eqnarray} 
\displaystyle\partial_{\zeta}(G+H)&=&\frac{\alpha_{t}}{2\sqrt{2}\alpha}
(G+H)
\nonumber\\ 
\displaystyle & &-\frac{1}{\sqrt{2}t}\left(U_{t}^2+\sqrt{\alpha}U_{t}U_{\theta}
+\frac{\mbox{e}^{4U}}{4t^2}(\alpha A_{\theta}^2+\sqrt{\alpha}A_{t}A_{\theta})\right)
\nonumber\\ 
\displaystyle & &+(U_{t}+\sqrt{\alpha}U_{\theta})\frac{\alpha}{2\sqrt{2}}
\mbox{e}^{2(\eta-U)}\kappa+\frac{\alpha\mbox{e}^{2\eta}}{2\sqrt{2}t}
(A_{t}+\sqrt{\alpha}A_{\theta})S_{23}, 
\nonumber\\ 
\label{han11}\\ 
\displaystyle\partial_{\chi}(G-H)&=&\frac{\alpha_{t}}{2\sqrt{2}\alpha}
(G-H)
\nonumber\\ 
\displaystyle & &-\frac{1}{\sqrt{2}t}\left(U_{t}^2-\sqrt{\alpha}U_{t}U_{\theta}
+\frac{\mbox{e}^{4U}}{4t^2}(\alpha A_{\theta}^2-\sqrt{\alpha}A_{t}A_{\theta})\right)
\nonumber\\ 
\displaystyle & &+(U_{t}-\sqrt{\alpha}U_{\theta})\frac{\alpha}{2\sqrt{2}}
\mbox{e}^{2(\eta-U)}\kappa+\frac{\alpha\mbox{e}^{2\eta}}{2\sqrt{2}t}
(A_{t}-\sqrt{\alpha}A_{\theta})S_{23}.
\nonumber\\ 
\label{hanvi} 
\end{eqnarray} 
Here $\kappa=\rho-P_1+P_2-P_3.$ 
Now we wish to integrate these equations along the integral curves of 
the vector fields $\partial_{\chi}$ and $\partial_{\zeta}$ respectively 
(let us henceforth call these integral curves null curves, since they 
are null with respect to the two-dimensional ``base spacetime''). Below 
we show that the quantity 
\begin{equation} 
\Gamma(t):=\sup_{\theta\in S^1}G(t,\cdot)+Q^2(t), 
\end{equation} 
is uniformly bounded on $[t_2,t_3)$ by deriving the inequality 
\begin{equation} 
\Gamma(t)\leq C+\int_{t_2}^{t}\Gamma(s)\ln{\Gamma(s)}ds.\label{GlogG} 
\end{equation} 
We begin with two observations. 
Let $\gamma$ and $X$ be a geodesic and a Killing vector field respectively in 
any spacetime. Then $g(\gamma',X)$ is conserved along the geodesic. Here 
$\gamma'$ is the tangent vector to $\gamma$. In our case we have the two 
Killing vector fields $\partial_{x}$ and $\partial_{y}$. The particles follow 
the geodesics of spacetime with tangent $p^{\mu}$, so $g_{\mu\nu}p^{\mu}
(\partial_{x})^{\nu}$ 
and $g_{\mu\nu}p^{\mu}(\partial_{y})^{\nu}$ are thus conserved. 
Expressing $p^{\mu}$ in 
terms of $v^{\mu}$ (see (\ref{pmu})) we find that 
$$V^{2}(t)\mbox{e}^{U(t,\Theta(t))}$$ and 
$$V^{2}(t)A\mbox{e}^{U}+V^{3}(t)t\mbox{e}^{-U(t,\Theta(t))},$$ 
are conserved. Here $V^{2}(t), V^{3}(t)$ and $\Theta(t)$ are solutions 
to the characteristic system associated to the Vlasov equation. From step 2 
we have that $U$ and $A$ are uniformly bounded on $[t_2,t_3)$. 
Hence $|V^2(t)|$ 
and $|V^3(t)|$ are both uniformly bounded on $[t_2,t_3)$, and since 
the initial distribution function $f_0$ has compact support we conclude that 
\begin{equation} 
\sup\{|v^2|+|v^3|:\exists (s,\theta)
\in [t_2,t]\times S^1\mbox{ with }f(s,\theta,v)\not= 0\}, 
\end{equation} 
is uniformly bounded on $[t_2,t_3)$. Therefore, in order to control 
$Q(t)$ it is sufficient to control 
\begin{equation} 
Q^1(t):=\mbox{sup}\{|v^1|:\exists (s,\theta)\in [t_2,t]\times 
S^1\mbox{such that} f(s,\theta,v)\not= 0\}.\label{Q1} 
\end{equation} 
Below we introduce the uniformly bounded function $\gamma(t)$ to denote 
estimates regarding 
the variables $v^2$ and $v^3$. 
Next we observe that there is some cancellation to take advantage of 
in the matter term $(\rho-P_1)$ which appears in 
the equations for $G+H$ and $G-H$ above. 
This term can be estimated as follows 
\begin{eqnarray} 
\displaystyle 0\leq(\rho-P_1)(t,\theta)&=&
\int_{\mathbb{R}^3}(v^0-\frac{(v^1)^2}{v^0})f(t,\theta,v)dv
\nonumber\\ 
\displaystyle &=&\int_{\mathbb{R}^3}\frac{1+(v^2)^2+(v^3)^2}{v^0}f(t,\theta,v)dv
\nonumber\\ 
\displaystyle &\leq& 
\int_{\mathbb{R}^3}[1+(v^2)^2+(v^3)^2]|f|\frac{dv}{\sqrt{1+(v^1)^2}}
\nonumber\\ 
\displaystyle &\leq&\|f_0\|_{\infty}\gamma(t)
\int_{|v^1|\leq Q^1(t)}\frac{dv^1}{\sqrt{1+(v^1)^2}}
\nonumber\\ 
\displaystyle &\leq& C\gamma(t)\ln{Q^1(t)}.\label{lnrp1} 
\end{eqnarray} 
In a similar fashion we can estimate $P_2,P_3$ and $S_{23}$. 
Indeed, for $k=1,2$, 
we have 
\begin{eqnarray} 
\displaystyle 0\leq P_k(t,\theta)&=&
\int_{\mathbb{R}^3}\frac{(v^k)^2}{v^0}f(t,\theta,v)dv 
\nonumber\\ 
\displaystyle&\leq&\|f_0\|_{\infty}\gamma(t)
\int_{|v^1|\leq Q^1(t)}\frac{dv^1}{\sqrt{1+(v^1)^2}}
\nonumber\\ 
\displaystyle &\leq& C\gamma(t)\ln{Q^1(t)}.\label{lnp23} 
\end{eqnarray} 
The argument is almost identical for $S_{23}$. 

\textbf{Remark. }Since the matter of interest is large momenta we 
have here assumed that $Q^1(t)\geq 2$ to 
avoid the introduction of some 
immaterial constants in the estimates.\\ 

Let us now derive (\ref{GlogG}). As in step 2 in section 4 we integrate 
the equations above for $G+H$ and $G-H$ along null paths. 
For $t\geq t_{2}$, let 
$$A(t,\theta)=\int_{t_2}^{t}\sqrt{\alpha}(s,\theta)ds,$$ 
and integrate along the two null paths defined by $\partial_{\chi}$ and 
$\partial_{\zeta}$, starting 
at $(t_2,\theta)$ and 
add the results. We get for $t\in [t_2,t_3)$, 
\begin{eqnarray} 
&\displaystyle G(t,\theta)=\frac{1}{2}[G+H](t_2,\theta-(A(t)-t_2))
+\frac{1}{2}[G+H](t_2,
\theta+(A(t)-t_2))&\nonumber\\ 
&\displaystyle +\frac{1}{2}\int_{t_2}^{t} 
K_{1}(s,\theta-(A(s)-t_2))+K_{2}(s,\theta+(A(s)-t_2))\,ds&\nonumber\\&
\displaystyle +\frac{1}{2}\int_{t_2}^{t} 
L_{1}(s,\theta-(A(s)-t_2))+L_{2}(s,\theta+(A(s)-t_2))\,ds&\nonumber\\ 
&\displaystyle +\frac{1}{2}\int_{t_2}^{t}
[U_{\chi}M](s,\theta-(A(s)-t_2))+ 
[U_{\zeta}M](s,\theta+(A(s)-t_2))\,ds&
\nonumber\\
&\displaystyle +\frac{1}{2}\int_{t_2}^{t}
[\frac{A_{\chi}}{2t}\tilde{M}](s,\theta-(A(s)-t_2))
+[\frac{A_{\zeta}}{2t}\tilde{M}](s,\theta+(A(s)-t_2))\,ds,&
\nonumber\\ 
\label{gphil} 
\end{eqnarray} 
where 
\begin{eqnarray} 
&\displaystyle K_{1}=\frac{\alpha_{t}}{2\sqrt{2}\alpha}
(G+H),\;\; K_{2}=\frac{\alpha_{t}}{2\sqrt{2}\alpha}
(G-H),&\\ 
&\displaystyle L_{1}=-\frac{1}{\sqrt{2}t}
\left(U_{t}^2+\sqrt{\alpha}U_{t}U_{\theta}+\frac{\mbox{e}^{4U}}{4t^2}
(\alpha A_{\theta}^2+\sqrt{\alpha}A_{t}A_{\theta})\right),&
\\
&\displaystyle L_{2}=-\frac{1}{\sqrt{2}t}
\left(U_{t}^2-\sqrt{\alpha}U_{t}U_{\theta}
+\frac{\mbox{e}^{4U}}{4t^2}(\alpha A_{\theta}^2-\sqrt{\alpha}A_{t}A_{\theta})\right),&\\ 
&\displaystyle M=\frac{1}{2}\mbox{e}^{2(\tilde{\eta}-U)}
\kappa,\;\; \tilde{M}=\mbox{e}^{2\tilde{\eta}}S_{23}.& 
\end{eqnarray} 
Note that in the expression for $M$ and $\tilde{M}$ we used 
$\alpha \mbox{e}^{2\eta}=\mbox{e}^{2\tilde{\eta}}$. 
It is easy to see that both $G+H$ and $G-H$ can be written as sums of two 
squares. 
From the constraint equation equation (\ref{constr3}) we find that 
$\alpha_{t}/\alpha\leq 0$ so that $K_1$ and $K_2$ are nonpositive. 
Using the elementary inequality $2ab\leq a^2+b^2$ and the fact that  
$|\tilde{\eta}|$ and $|U|$ are uniformly bounded we 
obtain from (\ref{gphil}) the inequality 
\begin{eqnarray}
\displaystyle\sup_{\theta}G(t,\cdot)&\leq&\sup_{\theta}
G(t_2,\cdot)+\sup_{\theta}H(t_2,\cdot)
+ C\int_{t_2}^{t}\frac{1}{s}\sup_{\theta}G(s,\cdot)ds
\nonumber\\ 
\displaystyle &+&\int_{t_2}^t C(s)\sup_{\theta}[\sqrt{G(s,\cdot)}
((\rho-P_1+P_2-P_3)+S_{23})]ds 
\nonumber\\ 
\displaystyle &\leq&C+C(t)\int_{t_2}^{t}[\sup_{\theta}G(s,\cdot)
+\sup_{\theta}\sqrt{G(s,\cdot)}\ln{Q^1(s)}]ds,\nonumber\\ 
\label{supg} 
\end{eqnarray} 
where (\ref{lnrp1}) and (\ref{lnp23}) were used in the last inequality. 

\textbf{Remark. }The sign of $K_1$ and $K_2$ simplified the estimate above. 
This is not crucial since $|\alpha_{t}|/\alpha$ is bounded 
by $\ln{Q^1(t)}$ which is sufficient for obtaining a bound on $\Gamma(t)$.\\ 

Let us now derive an estimate for $Q^1$ in terms of $\sup_{\theta}G$. 
\begin{lemma} 
Let $Q^1(t)$ and $G(t,\theta)$ be as above. Then 
\begin{equation} 
\displaystyle |Q^1(t)|^2\leq C+
D(t)\int_{t_2}^{t}[(Q^1(s))^2+\sup_{\theta}G(s,\cdot)]ds, 
\end{equation} 
where $C$ is a constant and $D(t)$ is a uniformly bounded function 
on $[t_2,t_3)$. 
\end{lemma} 
\textit{Proof. }The characteristic 
equation for $V^1$ associated to the Vlasov equation reads 
\begin{eqnarray} 
\displaystyle\frac{dV^1(s)}{ds}&=&-(\eta_{\theta}-U_{\theta}+
\frac{\alpha_{\theta}}{2\alpha})\sqrt{\alpha}V^{0}-(\eta_{t}-U_{t})V^{1} 
\nonumber\\ 
\displaystyle & &-\frac{\sqrt{\alpha}U_{\theta}}{V^{0}}
((V^{2})^{2}-(V^{3})^{2})
+\frac{\sqrt{\alpha}A_{\theta}}{sv^0}\mbox{e}^{2U}v^2v^3.\label{euhat} 
\end{eqnarray} 
We will now split the right hand side into three terms to be analyzed 
separately. 
Expressing $\eta_{\theta}$ and $\eta_{t}$ by 
using the constraint equations (\ref{constr1}) and (\ref{constr2}) we obtain 
\begin{equation} 
\frac{d}{ds}(V^1(s))^2=2V^1(s)\frac{d}{ds}V^1(s)=T_1+T_2+T_3, 
\end{equation} 
where 
\begin{eqnarray}
\displaystyle T_{1}&=&-2V^1(s)[s\alpha\mbox{e}^{2(\eta-U)}(JV^0+\rho V^1)], 
\nonumber\\ 
\displaystyle 
T_{2}&=&-2V^1(s)\left[s(U_{t}^2+\alpha U_{\theta}^2+\frac{\mbox{e}^{4U}}{4s^2}
(A_{t}^2+\alpha A_{\theta}^2))V^1\right. 
\nonumber\\ 
\displaystyle & &\left.+2s\sqrt{\alpha}
U_{\theta}U_{t}V^0-\sqrt{\alpha}U_{\theta}V^0-U_{t}V^1+\frac{\mbox{e}^{4U}}{2s} 
\sqrt{\alpha}A_{t}A_{\theta}V^0\right], 
\nonumber\\ 
\displaystyle T_{3}&=&-2V^1(s)[\frac{\sqrt{\alpha}U_{\theta}}{V^0}
((V^3)^2-(V^2)^2)-\frac{\sqrt{\alpha}A_{\theta}}{sV^0}\mbox{e}^{2U}V^2V^3].
\nonumber 
\end{eqnarray} 
Let us first estimate $T_1$. We split it into two terms 
\begin{equation} 
T_{1}=T_{1}^{-}+T_{1}^{+}=-2sV^1(s) 
\mbox{e}^{2(\tilde{\eta}-U)}(I^{-}+I^{+}),
\end{equation} 
where 
\begin{eqnarray} 
\displaystyle I^{-}&=&\int_{\mathbb{R}^2}\int_{-\infty}^{0}(v^1V^0+v^0V^1)
f(s,\theta,v)dv^1dv^2dv^3, 
\nonumber\\ 
\displaystyle I^{+}&=&\int_{\mathbb{R}^2}\int_{0}^{\infty}(v^1V^0+v^0V^1)
f(s,\theta,v)dv^1dv^2dv^3. 
\nonumber 
\end{eqnarray} 
Let us now consider the two cases $V^1(s)>0$ and $V^1(s)<0$. On a time 
interval where $V^1(s)>0$, $I^{+}$ is nonnegative and $T_{1}^{+}$ 
can therefore be discarded since it is nonpositive. The kernel 
in $I1^{-}$ can be estimated as follows 
\begin{eqnarray} 
\displaystyle  
v^1V^0+v^0V^1&=&\frac{(v^1)^2(V^0)^2-(v^0)^2(V^1)^2}{v^1V^0-v^0V^1}
\nonumber\\ 
\displaystyle &=&\frac{(v^1)^2(1+(V^2)^2+(V^3)^2)}
{v^1V^0-v^0V^1}+\frac{(V^1)^2(1+(v^2)^2+(v^3)^2)}{v^0V^1-v^1V^0}. 
\nonumber 
\end{eqnarray} 
Of course, the cancellation of the terms $(v^1)^2(V^1)^2$ is essential 
in this compuation. 
The second term is positive since $V^1(s)>0$ and $v^1<0$, 
and contributes negatively to $T_{1}^{-}$ and can be discarded. The first 
term is negative and the modulus can be estimated by 
\begin{equation} 
\frac{(v^1)^2(1+(V^2)^2+(V^3)^2)}{|v^1|V^0+v^0V^1}\leq\frac{|v^1|
(1+(V^2)^2+(V^3)^2)}{V^1}. 
\end{equation} 
In the expression for $T_{1}$ we first note that 
$2s\alpha\mbox{e}^{2(\eta-U)}=2s\mbox{e}^{2(\tilde{\eta}-U)}\leq C(s)$. 
Hence, on the time interval where $V^1(s)>0$ we can 
estimate $T_{1}$ by 
\begin{eqnarray} 
\displaystyle T_{1}&\leq&T_{1}^{-}\leq \|f_0\|_{\infty}C(s)V^1(s)\int_{\mathbb{R}^2}
\int_{0}^{Q^1}\frac{v^1
(1+(V^2)^2+(V^3)^2)}{V^1(s)}dv^1du
\nonumber\\ 
\displaystyle &\leq&\|f_0\|C(s)\gamma(s)\int_{0}^{Q^1}v^1dv^1 
\leq C(s)(Q^1(s))^2. 
\end{eqnarray} 
On a time interval where $V^1<0$ we see that $T_{1}^{-}$ is nonpositive and 
can be discarded. We can then estimate $T_{1}^{+}$ by using almost identical 
arguments as for $T_{1}^{-}$ and we get also on such a time interval, 
\begin{equation} 
T_{1}\leq T_{1}^{+}\leq C(s)(Q^1(s))^2. 
\end{equation} 
Let us now consider $T_{2}$. 
We again study the cases $V^1(s)>0$ and $V^1(s)<0$. 
Assume first that $V^1(s)>0$ on some time interval. 
The expression for $T_{2}$ can be 
written $T_{2}=T_{2}^{p}+T_{2}^{r}$ (p=principal, r=rest) where 
\begin{displaymath} 
T_{2}^{p}=-2(V^1(s))^2\left(
[s(U_{t}+\sqrt{\alpha}U_{\theta})^{2}-(U_{t}+\sqrt{\alpha}U_{\theta})]
+[\frac{\mbox{e}^{4U}}{4s}(A_{t}+\sqrt{\alpha}A_{\theta})^2]\right) 
\end{displaymath} 
and 
\begin{displaymath} 
T_{2}^{r}=2(V^0(s)-V^1(s))V^1(s)[\sqrt{\alpha}
U_{\theta}-2s\sqrt{\alpha}U_{t}U_{\theta}-\frac{\mbox{e}^{4U}}{2s}\sqrt{\alpha}
A_{t}A_{\theta}]. 
\end{displaymath} 
For $T_{2}^{r}$ we have 
\begin{eqnarray} 
\displaystyle |T_{2}^{r}|&=&\frac{2(1+(V^2)^2+(V^3)^2)V^1(s)}{V^0+V^1}
\sqrt{\alpha}|U_{\theta}-2sU_{t}U_{\theta}-\frac{\mbox{e}^{4U}}{2s}
A_{t}A_{\theta}|
\nonumber\\ 
\displaystyle &\leq&(s+1)\gamma(s)\sup_{\theta}G(s,\cdot). 
\end{eqnarray} 
Since the matter of interest is large $G$ we have here 
assumed that $\sqrt{G}\leq G$. This assumption will be used below without 
comment. 
To estimate $T_{2}^{p}$ we observe that for $s\geq t_2$, 
$$sa^2-a\geq \frac{-1}{4s}\geq\frac{-1}{4t_{2}},\mbox{ for any }a\in R.$$ 
The term involving $A$ contributes negatively and can be discarded, thus 
\begin{equation} 
T_{2}^{p}\leq\frac{1}{2t_{2}}(V^1(s))^2\leq C(Q^1(s))^2. 
\end{equation} 
On a time interval where $V^1(s)<0$, the same estimates hold. 
Indeed, we only have to 
write $T_{2}=T_{2}^{p}+T_{2}^{r}$ in the form 
\begin{displaymath} 
T_{2}^{p}=-2(V^1(s))^2
\left([s(U_{t}-\sqrt{\alpha}U_{\theta})^{2}- (U_{t}-\sqrt{\alpha}U_{\theta})]
+\frac{\mbox{e}^{4U}}{4s}(A_{t}-\sqrt{\alpha}A_{\theta})^2\right) 
\end{displaymath} 
and 
\begin{displaymath} 
T_{2}^{r}=2(V^0(s)+V^1(s))V^1(s)
[\sqrt{\alpha}U_{\theta}-2s\sqrt{\alpha}U_{t}U_{\theta}-\frac{\mbox{e}^{4U}}{2s}
\sqrt{\alpha}A_{t}A_{\theta}], 
\end{displaymath} 
and the same arguments apply. 
Therefore we have obtained 
\begin{equation} 
T_{2}\leq T_{2}^{p}+|T_{2}^{r}|\leq C(Q^1(s))^2+C(s)\sup_{\theta}G(s,\cdot). 
\end{equation} 
Finally we estimate $T_{3}$. It follows immediately that 
\begin{equation} 
|T_{3}|\leq\gamma(s)\frac{|V^1(s)|}{V^0}\sqrt{\alpha}|U_{\theta}
+\frac{\mbox{e}^{2U}}{s}A_{\theta}| 
\leq C(s)\sup_{\theta}G(s,\cdot). 
\end{equation} 
The lemma now follows by adding the estimates for $T_k$, $k=1,2,3$. 
\begin{flushright} 
$\Box$ 
\end{flushright} 
Combining the estimate for $(Q^1(t))^2$ in the lemma and the estimate 
(\ref{supg}) for $\sup_{\theta}G(t,\cdot)$, we find that $\Gamma(t)$ 
satisfies the estimate (\ref{GlogG}) and is thus uniformly bounded. 
The constraint equation (\ref{constr1}) now immediately shows that 
$|\eta_{t}|$ is bounded by $$2tG+t\mbox{e}^{2(\tilde{\eta}-U)}\rho
\leq C(t)[\sup_{\theta}G(t,\cdot)+(Q(t))^3],$$ 
since $$\rho=\int_{\mathbb{R}^3}fdv\leq
\|f_0\|_{\infty}\int_{|v|\leq Q(t)}dv\leq C(Q(t))^3.$$ 
Analogous arguments show that $|\alpha_{t}|$ is uniformly bounded. 
The uniform bound on $G$ provides bounds on $|U_t|$ and $|A_{t}|$, but 
to conclude that $|U_{\theta}|$ and $|A_{\theta}|$ are bounded we have to 
show that $\alpha$ 
stays uniformly bounded away from zero. Equation (\ref{constr3}) is easily 
solved, 
\begin{equation} 
\displaystyle \alpha(t,\theta)=\alpha(t_2,\theta)
\mbox{e}^{\int_{t_2}^tF(s,\theta)ds}, 
\end{equation} 
where $$F(t,\theta):=-2t\mbox{e}^{2(\tilde{\eta}-U)}(\rho-P_1),$$ 
which is uniformly bounded from below. 
Hence $|U_{\theta}|$ and $|A_{\theta}|$ are bounded 
and step 2 is complete. 

\textit{Step 3. }(Bounds on $\partial f$, $\alpha_{\theta}$ 
and $\eta_{\theta}$.)\\ 
The main goal in this step is to show that the first derivatives 
of the distribution function are 
bounded. In view of the bound on $Q(t)$ we then also obtain bounds 
on the first derivatives of the matter terms $\rho, J, S_{23}$ and 
$P_k,$ $k=1,2,3$. 
Such bounds almost immediately lead to bounds on $\alpha_{\theta}$ 
and $\eta_{\theta}$. 

Recall that the solution $f$ can be written in the form 
\begin{equation} 
f(t,\theta,v)=f_{0}(\Theta(0,t,\theta,v),V(0,t,\theta,v)),\label{solu1} 
\end{equation} 
where $\Theta(s,t,\theta,v), V(s,t,\theta,v)$ is the solution 
to the characteristic system
\begin{eqnarray} 
\displaystyle\frac{d\Theta}{ds}&=&\sqrt{\alpha}\frac{V^1}{V^0}, 
\label{chtta}\\ 
\displaystyle\frac{dV^1}{ds}&=&-
(\eta_{\theta}-U_{\theta}+\frac{\alpha_{\theta}}{2\alpha})\sqrt{\alpha}V^{0}-(\eta_{t}-U_{t})V^{1}
\nonumber\\
\displaystyle & &-\sqrt{\alpha}U_{\theta}
\frac{(V^3)^2- (V^{2})^{2}}{V^{0}}+\frac{\mbox{e}^{2U}}{s}\sqrt{\alpha}A_{\theta}
\frac{V^2V^3}{V^0}, 
\label{chdv1}\\ 
\displaystyle\frac{dV^2}{ds}&=&-U_{t}V^{2}-\sqrt{\alpha}U_{\theta}
\frac{V^{1}V^{2}}{V^{0}},
\label{chdv2}\\ 
\displaystyle\frac{dV^3}{ds}&=&-(\frac{1}{s}-U_{t})V^{3}
+\sqrt{\alpha}U_{\theta}\frac{V^{1}V^{3}}{V^{0}}
\nonumber\\ 
\displaystyle & &-\frac{\mbox{e}^{2U}}{s}(A_{t}+\sqrt{\alpha}A_{\theta}\frac{V^1}{V^0})V^2,
\label{chdv3} 
\end{eqnarray} 
with the property 
$\Theta(t,t,\theta,v)=\theta$, $V(t,t,\theta,v)=v$. 
Hence, in order to establish bounds on the first derivatives of $f$ it is 
sufficient to bound $\partial\Theta$ and $\partial V$ since $f_0$ is smooth. 
Here $\partial$ denotes the first order derivative with respect 
to $t,\theta$ or $v$. 
Evolution equations for $\partial\Theta$ and $\partial V$ are provided 
by the characteristic system above. However, the right hand sides will 
contain second order derivatives of the field components, but so far 
we have only obtained bounds on the first order derivatives 
(except for $\eta_{\theta}, \alpha_{\theta}$). 
Yet, certain combinations of second order derivatives can 
be controlled. Behind this observation lies 
a geometrical idea which plays a fundamental role in 
general relativity. An important property of curvature is its control 
over the relative behaviour of nearby geodesics. Let $\gamma(u,\lambda)$ be 
a two-parameter family of geodesics, i.e. for each fixed $\lambda$, 
the curve $u\mapsto\gamma(u,\lambda)$ is a geodesic. 
Define the variation vector field 
$Y:=\gamma_{\lambda}(u,0)$. This vector field satisfies 
the geodesic deviation equation (or Jacobi equation) (see eg. [HE]) 
\begin{equation} 
\frac{D^2Y}{Du^2}=R_{Y\gamma'}\gamma',\label{ged} 
\end{equation} 
where $D/Du$ is the covariant derivative, $R$ the Riemann curvature tensor, 
and $\gamma':=\gamma_{u}(u,0)$. 
Now, the Einstein tensor is closely related to the curvature tensor and 
since the Einstein tensor is proportional to the energy momentum tensor 
which we can control from step 2, it is meaningful, in view of (\ref{ged}) 
(with $Y=\partial \Theta$), 
to look for linear combinations of $\partial\Theta$ and $\partial V$ which 
satisfy an equation with bounded coefficients. More precisely, we want to 
substitute the twice differentiated field components which appear by 
taking the derivative of the characteristic system by using 
the Einstein equations. 
The geodesic deviation equation has previously played an important role 
in studies of the Einstein-Vlasov system ([RR], [Rn] and [Rl3]). 
\begin{lemma} 
Let $\Theta(s)=\Theta(s,t,\theta,v)$ and $V^k(s)=V^k(s,t,\theta,v)$, 
$k=1,2,3$ be a solution to the characteristic system 
(\ref{chtta})-(\ref{chdv3}). Let $\partial$ denote 
$\partial_{t},\partial_{\theta}$ 
or $\partial_{v}$, and define 
\begin{eqnarray} 
\displaystyle\Psi&=&\alpha^{-1/2}\partial\Theta,\label{Psi}\\ 
\displaystyle Z^1&=&\partial V^1+
\left(\frac{\eta_tV^0}{\sqrt{\alpha}}-\frac{U_tV^0}
{\sqrt{\alpha}}\,\frac{(V^0)^2-(V^1)^2+(V^2)^2-(V^3)^2}{(V^0)^2-(V^1)^2}
\right.\nonumber\\ 
\displaystyle & &\left.+\,U_{\theta}\,
\frac{V^1((V^2)^2-(V^3)^2)}{(V^0)^2-(V^1)^2}-\frac{A_{t}\mbox{e}^{2U}}
{\sqrt{\alpha}t}\frac{V^0V^2V^3}{(V^0)^2-(V^1)^2}\right. 
\nonumber\\ 
\displaystyle & & \left.+A_{\theta}
\frac{V^1V^2V^3}{(V^0)^2-(V^1)^2}\right)\partial\Theta,
\label{Z1}\\ 
\displaystyle Z^2&=&\partial V^2+V^2U_{\theta}\,\partial\Theta,
\label{Z2}\\ 
\displaystyle Z^3&=& 
\partial V^3-(V^3U_{\theta}-\frac{\mbox{e}^{2U}}{s}V^2A_{\theta})\,
\partial\Theta.\label{Z3} 
\end{eqnarray} 
Then there is a matrix $A=\{a_{lm}\}$, $l,m=0,1,2,3$, 
such that $$\Omega:=(\Psi,Z^1,Z^2,Z^3)^T$$ satisfies 
\begin{equation} 
\displaystyle\frac{d\Omega}{ds}=A\Omega,\label{Omega} 
\end{equation} 
and the matrix elements $a_{lm}=a_{lm}(s,\Theta(s),V^k(s))$ are 
all uniformly bounded on $[t_2,t_3)$. 
\end{lemma} 
\textit{Sketch of proof. }Once the ansatz 
(\ref{Psi})-(\ref{Z3}) has been found this is only a lengthy calculation. 
To illustrate the type of calculations 
involved we show the easiest case, i.e. the $Z^2$ term. 
\begin{eqnarray} 
\displaystyle\frac{dZ^2}{ds}&=&\frac{d}{ds}(\partial V^2+V^2U_{\theta}
\partial\Theta)\nonumber\\
\displaystyle&=&\partial(\frac{d}{ds}V^2)
+\frac{dV^2}{ds}U_{\theta}\partial\Theta\nonumber\\ 
\displaystyle&\phantom{h}&+V^2(U_{t\theta}+U_{\theta\theta}\frac{d\Theta}{ds})
\partial\Theta+V^2U_{\theta}\partial(\frac{d\Theta}{ds}).\nonumber\\ 
\end{eqnarray} 
Now we use (\ref{chtta}) and (\ref{chdv2}) to substitute for $d\Theta/ds$ 
and $dV^2/ds$. We find that the right hand side equals 
\begin{eqnarray} 
&\displaystyle\partial(-U_tV^2-\sqrt{\alpha}U_{\theta}\frac{V^1V^2}{V^0})+
(-U_tV^2-\sqrt{\alpha}U_{\theta}\frac{V^1V^2}{V^0})U_{\theta}\partial\Theta&
\nonumber\\ 
&\displaystyle +V^2(U_{t\theta}+U_{\theta\theta}\sqrt{\alpha}\frac{V^1}{V^0})
\partial\Theta+V^2U_{\theta}\left(\frac{\alpha_{\theta}V^1}{2\sqrt{\alpha}V^0}
\partial\Theta+\sqrt{\alpha}\partial(\frac{V^1}{V^0})\right)&. 
\nonumber
\end{eqnarray} 
Taking the $\partial$ derivative of the first term we find that all 
terms of second order derivatives and terms containing $\alpha_{\theta}$ 
cancel. Next, since 
\begin{equation} 
\displaystyle -\sqrt{\alpha}U_{\theta}\partial\left(\frac{V^1V^2}{V^0}\right)
+\sqrt{\alpha}U_{\theta}V^2\partial\left(\frac{V^1}{V^0}\right)=-\sqrt{\alpha}
U_{\theta}\frac{V^1}{V^0}\partial V^2, 
\end{equation} 
we are left with 
\begin{equation} 
\displaystyle \frac{dZ^2}{ds}=-(U_tV^2+\sqrt{\alpha}
U_{\theta}\frac{V^1V^2}{V^0})U_{\theta}\partial\Theta-(U_t+\sqrt{\alpha}
U_{\theta}\frac{V^1}{V^0})\partial V^2.\label{z2b} 
\end{equation} 
Finally we express this in terms of $\Psi, Z^1,Z^2$ and $Z^3$. Here 
this is easy and we immediately get 
\begin{displaymath} 
\frac{dZ^2}{ds}=-(U_{t}+\sqrt{\alpha}U_{\theta}\frac{V^1}{V^0})Z^2. 
\end{displaymath} 
Clearly, the map $(\partial\Theta,\partial V^k)\mapsto (\Psi,Z^k)$ is 
invertible so that this step is easy also in the other cases. 
It follows that the matrix elements $a_{2m}$, $m=0,1,2,3$, are 
uniformly bounded on $[t_2,t_3)$ (only $a_{22}$ is nonzero here). 
The computations for the other terms are similar. For the $Z^1$ term we point 
out that the evolution equations (\ref{evol2}) and (\ref{evol4}) should be 
invoked and that the matrix element $a_{10}$ contains $\eta_{\theta}$ 
and $\alpha_{\theta}/2\alpha$, but they combine and form $\tilde{\eta}$. 
\begin{flushright} 
$\Box$ 
\end{flushright} 
From the lemma it now immediately follows that $|\Omega|$ is 
uniformly bounded on $[t_2,t_3)$. Moreover, 
since the system (\ref{Psi})-(\ref{Z3}) is invertible with 
uniformly bounded coefficients 
we also have uniform bounds 
on $|\partial\Theta|$ and $|\partial V^k|$, $k=1,2,3$. 
In view 
of the discussion at the beginning of this section we see that 
the distribution function $f$ and the matter quantities 
$\rho, J, S_{23}$ and $P_k$, are all uniformly $C^1$ bounded. 
From the constraint equation (\ref{constr3}) we now obtain a uniform 
bound on $\alpha_{\theta}$ by a simple Gronwall argument using as usual 
the identity $\alpha\mbox{e}^{2(\eta-U)}=\mbox{e}^{2(\tilde{\eta}-U)}$. 
Finally this yields a uniform bound on $\eta_{\theta}$ since 
$$\eta_{\theta}=\tilde{\eta}_{\theta}-\frac{\alpha_{\theta}}{2\alpha}$$ 
and $\alpha$ stays uniformly bounded away from zero. 

\textit{Step 4. }(Bounds on second and higher 
order derivatives.)\\ 
It is now easy to obtain bounds on second order derivatives on $U$ and $A$ 
by using light cone arguments. We define $G$ and $H$ by 

\begin{eqnarray} 
G&=&\frac{1}{2}(U_{tt}^2+\alpha U_{t\theta}^2)+\frac{\mbox{e}^{4U}}{8t^2}
(A_{tt}^2+\alpha A_{t\theta}^2),\label{GEt}\\
H&=&\sqrt{\alpha}U_{tt}U_{t\theta}+\frac{\mbox{e}^{4U}}{4t^2}A_{tt}A_{t\theta},
\label{HEt} 
\end{eqnarray} 
and use the differentiated (with respect to $t$) evolution equations for $U$ 
and $A$ to obtain equations similar to (\ref{han11}) and (\ref{hanvi}). 
In this case a straightforward light cone argument applies since we have 
control of the differentiated matter terms. $U_{\theta\theta}$ 
and $A_{\theta\theta}$ are then 
uniformly bounded in view of the evolution equations 
(\ref{evol2}) and (\ref{evol4}). 
Bounds on second order derivatives on $f$ then follows from (\ref{Omega}) by 
studying the equation for $\partial\Omega$. The only thing to notice is 
that $\tilde{\eta}_{\theta\theta}$ is controlled by (\ref{constr2}). 
It is clear that this reasoning can be continued to give uniform bounds 
on $[t_2,t_3)$ for higher order derivatives as well. In view 
of the discussion after the statement of Theorem 1 in section 3, 
this completes the proof of Theorem 1 in the expanding direction. 
\begin{flushright} 
$\Box$ 
\end{flushright} 

\phantom{hej} 

\begin{center}
\textbf{Acknowledgement} 
\end{center} 
I am most grateful to Alan Rendall for suggesting the problem (for small data) 
and for commenting on the manuscript. This work was supported by the Swedish 
Foundation for International Cooperation in Research and Higher 
Education (STINT) and is hereby gratefully acknowledged.



\begin{thebibliography}{AAAA}

\bibitem[BCIM]{b1}{\sc B.K. Berger, P. Chru\'{s}ciel, J. Isenberg and 
V. Moncrief, }Global foliations of vacuum spacetimes with $T^2$ isometry, 
{\em Ann. Phys. }260, 117-148 (1997). 
\bibitem[BT]{b2}{\sc J. Binney and S. Tremaine, }{\em Galactic dynamics, }
Princeton University Press, Princeton (1987). 
\bibitem[CB]{c1}{\sc Y. Choquet-Bruhat, }Probl\`{e}me de Cauchy pour le 
syst\`{e}me int\'{e}gro diff\'{e}rentiel d'Einstein-Liouville, 
{\em Ann. Inst. Fourier }21, 181-201 (1971). 
\bibitem[CBG]{c2}{\sc Y. Choquet-Bruhat and R. Geroch, }Global aspects 
of the Cauchy problem in general relativity, {\em Comm. Math. Phys. }14, 
344-357 (1969). 
\bibitem[Cu1]{c3}{\sc D. Christodoulou, }Examples of naked singularity 
formation in the gravitational collapse of a scalar field, 
{\em Ann. Math. }140, 607-653 (1994). 
\bibitem[Cu2]{c4}{\sc D. Christodoulou, }Bounded variation solutions of 
the spherically symmetric Einstein-scalar field equations, 
{\em Comm. Pure Appl. Math. }46, 1131-1220 (1993). 
\bibitem[Cu3]{c7}{\sc D. Christodoulou, }Self-gravitating relativistic fluids: 
the formation of a free phase boundary in the phase transition from soft 
to hard, {\em Arch. Rational Mech. Anal. }134, 97-154 (1996). 
\bibitem[CK]{c8}{\sc D. Christodoulou and S. Klainerman, }
{\em The global nonlinear stability of the Minkowski space, }
Princeton University Press, Princeton 1993. 
\bibitem[Cl]{c9}{\sc P.T. Chru\'{s}ciel, }On spacetimes with $U(1)\times U(1)$ 
symmetric compact Cauchy surfaces, {\em Ann. Phys. }202, 100-150 (1990). 
\bibitem[CIM]{c10}{\sc P.T. Chru\'{s}ciel, J. Isenberg and V. Moncrief, }
Strong cosmic censorship in polarised Gowdy spacetimes, 
{\em Class. Quantum Grav. }7, 1671-1680 (1990). 
\bibitem[EM]{e1}{\sc D. Eardley and V. Moncrief, }The global existence 
problem and cosmic censorship in general relativity, {\em Gen. Rel. Grav. }
13, 887-892 (1981). 
\bibitem[ES]{m2}{\sc D. Eardley and L. Smarr, }Time functions in numerical 
relativity: marginally bound dust collapse, {\em Phys. Rev. }D19, 2239-2259 
(1979). 
\bibitem[E]{e2}{\sc J. Ehlers, }Survey of general relativity theory, 
In: W. Israel (ed.) {\em Relativity, Astrophysics and Cosmology. }
Reidel, Dordrecht 1973. 
\bibitem[GS]{g1}{\sc R. Glassey and W. Strauss, }Singularity formation in a
  collisionless plasma could only occur at high velocities, {\em
  Arch. Rat. Mech. Anal. } 92, 56-90 (1986).
\bibitem[G]{g3}{\sc R. Gowdy, }Vacuum spacetimes and compact invariant 
hypersurfaces: topologies and boundary conditions, 
{\em Ann. Phys. }83, 203-24 (1974). 
\bibitem[HE]{h1}{\sc S. Hawking and G. Ellis, }{\em The large scale structure 
of spacetime. }Cambridge University Press, Cambridge (1973). 
\bibitem[M]{m1}{\sc V. Moncrief }Global properties of Gowdy spacetimes with 
$T^3\times\mathbb{R}$ topology, {\em Ann. Phys. }132, 87-107 (1981). 
\bibitem[Rn]{r1}{\sc G. Rein }Cosmological solutions of the Vlasov-Einstein 
system with spherical, plane and hyperbolic symmetry, 
{\em Math. Proc. Camb. Phil. Soc. }119, 739-762 (1996). 
\bibitem[RR]{r2}{\sc G. Rein and A.D. Rendall, }Global existence of
  solutions of the spherically symmetric Vlasov-Einstein system with
  small initial data, {\em Comm. Math. Phys. }150, 561-583 (1992). Erratum: 
{\em Comm. Math. Phys. }176, 475-478 (1996). 
\bibitem[Rl1]{r3}{\sc A.D. Rendall, }Existence of constant mean curvature 
foliations in spacetimes with two-dimensional local symmetry, 
{\em Comm. Math. Phys. }189, 145-164 (1997). 
\bibitem[Rl2]{r4}{\sc A.D. Rendall, }Crushing singularities in spacetimes with 
spherical, plane and hyperbolic symmetry, {\em Class. Quantum Grav. }12, 
1517-1533 (1995). 
\bibitem[Rl3]{r5}{\sc A.D. Rendall, }An 
introduction to the Einstein-Vlasov system. Mathematics of gravitation, 
part I (Warsaw, 1996), 35-68, {\em Banach center Publ. }41 part I, 
Polish Acad. Sci., Warsaw (1997). 
\bibitem[Rl4]{r6}{\sc A.D. Rendall, }On the choice of matter model in general 
relativity, In: R. d'Inverno (ed.) {\em Approaches to Numerical Relativity. }
Cambridge University Press, Cambridge 1992. 
\bibitem[Rl5]{r7}{\sc A.D. Rendall, }Cosmic censorship 
and the Vlasov equation, {\em Class. Quantum Grav. }9, L99-L104 (1992). 
\bibitem[RRS]{r8}{\sc G. Rein, A.D Rendall and J. Schaeffer, }A regularity 
theorem for solutions of the spherically symmetric Vlasov-Einstein system, 
{\em Comm. Math. Phys. }168, 467-478 (1995).
\end{thebibliography}
\end{document}